\documentclass[aps,prb,reprint,groupedaddress]{revtex4-1}

\usepackage{braket}
\usepackage{amsmath}
\usepackage[colorinlistoftodos]{todonotes}
\usepackage[colorlinks,citecolor={blue}]{hyperref}% add hypertext capabilities
\usepackage{subcaption}
\usepackage{url}

\begin{document}

% Use the \preprint command to place your local institutional report
% number in the upper righthand corner of the title page in preprint mode.
% Multiple \preprint commands are allowed.
% Use the 'preprintnumbers' class option to override journal defaults
% to display numbers if necessary
%\preprint{}

%Title of paper
\title{Exciting determinants in Quantum Monte Carlo: Loading the dice with fast, low memory weights}

% repeat the \author .. \affiliation  etc. as needed
% \email, \thanks, \homepage, \altaffiliation all apply to the current
% author. Explanatory text should go in the []'s, actual e-mail
% address or url should go in the {}'s for \email and \homepage.
% Please use the appropriate macro foreach each type of information

% \affiliation command applies to all authors since the last
% \affiliation command. The \affiliation command should follow the
% other information
% \affiliation can be followed by \email, \homepage, \thanks as well.
\author{Verena A. Neufeld}
 \email{van26@cam.ac.uk}
\author{Alex J. W. Thom}%
 \email{ajwt3@cam.ac.uk}
\affiliation{%
 Department of Chemistry, Lensfield Road, Cambridge, CB2 1EW, United Kingdom
}%

%Collaboration name if desired (requires use of superscriptaddress
%option in \documentclass). \noaffiliation is required (may also be
%used with the \author command).
%\collaboration can be followed by \email, \homepage, \thanks as well.
%\collaboration{}
%\noaffiliation

\date{\today}

\begin{abstract}
High-quality excitation generators are crucial to the effectiveness of Coupled cluster Monte Carlo (CCMC) and full configuration interaction Quantum Monte Carlo (FCIQMC) calculations.
The heat bath sampling of Holmes et al. [A. A. Holmes, H. J. Changlani, and C. J. Umrigar, J.
Chem. Theory Comput. 12, 1561 (2016)] dramatically increases the efficiency
of the spawn step of such algorithms but requires memory storage scaling quartically with system size which can be prohibitive for large systems. Alavi {\em et al.} [S. D. Smart, G. H. Booth, and A. Alavi, unpublished] then approximated these weights with
weights based on Cauchy--Schwarz-like inequalities calculated on-the-fly. While reducing the memory cost, this algorithm scales linearly
in system size computationally. 
We combine both these ideas with the single reference nature of many systems, and introduce a spawn-sampling
algorithm that has low memory requirements (quadratic in basis set size) compared to the heat bath algorithm and only scales either independently of system size (CCMC)
or linearly in the number of electrons (FCIQMC). On small water chains with localized orbitals, we show that it is equally
efficient as the other excitation generators. As the system gets larger, it converges faster than the on-the-fly weight algorithm, while having a much more favourable memory scaling than the heat bath algorithm.
\end{abstract}

% insert suggested PACS numbers in braces on next line
% insert suggested keywords - APS authors don't need to do this
%\keywords{}

%\maketitle must follow title, authors, abstract, \pacs, and \keywords
\maketitle

\section{Introduction}
Coupled cluster Theory \cite{Coester1960,Cizek1966,Cizek1971,Bartlett2007} can 
give ground state energies to chemical accuracy (1 kcal mol$^{-1}$) \cite{Bartlett2007,Lee1995} in a  
systematically improvable manner. As an alternative to deterministically solving the coupled cluster equations, stochastic coupled cluster (CCMC) 
\cite{Thom2010,Spencer2016,Franklin2016,Scott2017b,Spencer2018} adopts a sparse representation of the wavefunction which can reduce memory requirements compared to a full deterministic representation. This
enables the use of higher coupled cluster levels and larger 
basis sets. Recently \cite{Neufeld2017b}, a finite uniform electron gas has been 
studied with truncation levels coupled cluster singles and doubles (CCSD) up to 
 quintuple excitations (CCSDTQ5) reaching basis set 
sizes of 18342 spinorbitals. 
Like full configuration interaction 
quantum Monte Carlo (FCIQMC) \cite{Booth2009,Cleland2010}, CCMC also does not suffer from the 
fermion sign problem in the same way\cite{Spencer2012} as diffusion Monte Carlo (DMC)\cite{Foulkes2001} does.
Provided there are enough walkers in the calculations\cite{Cleland2010,Spencer2012}, FCIQMC energies are exact
for the basis set chosen. FCIQMC has been applied to various molecules\cite{Booth2011, Cleland2012, Daday2012, Booth2014, Holmes2016a,Sharma2014,Veis2018,Samanta2018} and some periodic systems\cite{Spencer2012,Shepherd2012,Shepherd2012a,Shepherd2012b,Booth2013,Schwarz2015, Neufeld2017b,Luo2018a,Ruggeri2018} to find
ground state energies. It has also been used to determine excited state energies for example \cite{Booth2012,Ten-no2013,Humeniuk2014,Blunt2015,Blunt2017}. Both CCMC and FCIQMC have been used with the CC(P;Q) technique \cite{Deustua2017},
which can speed up the time needed to find the main excitors/determinants in CC(P;Q).
The algorithm used to perform FCIQMC and CCMC affects the speed
 and convergence, and there is still great scope for improvements\cite{Cleland2010,Petruzielo2012, Blunt2015b,Kersten2016, Holmes2016a,Smartunpub, Blunt2018, Spencer2016,Franklin2016,Scott2017b}.
Here, we propose a change to the \textit{spawn} step in the algorithm 
to use weighted excitations, inspired by the heat bath algorithm 
proposed by Holmes et al. \cite{Holmes2016a} (which was then expanded to the heat bath configuration interaction
algorithm\cite{Holmes2016,Sharma2017,Holmes2017}), and Cauchy--Schwarz weights proposed 
by Smart et al. \cite{Smartunpub}. The method introduced here has a lower computational scaling
than the heat bath excitation generators and a significantly lower memory cost.
\par The main part of Fock space quantum Monte Carlo algorithms such as
CCMC and FCIQMC consists of four steps;
\textit{selection} of a determinant/an excitor, \textit{spawn}, \textit{death} and \textit{annihilation}.
The \textit{spawn} part of the algorithm explores the space of possible
determinants/excitors. For a given determinant, it decides how the determinants connected via the hamiltonian  become
involved in the wavefunction by becoming occupied. As Holmes et al. \cite{Holmes2016a} already
noted, it is not efficient to give all determinants/excitors an equal probability of being
considered as some are more important for the dynamics than others. They have shown that
their heat bath weighting when selecting states to spawn to can greatly improve the overall efficiency.
\par A Slater determinant 
$\ket{\mathrm{D}_{\mathbf{m}}}$ is connected to another determinant 
$\ket{\mathrm{D}_{\mathbf{n}}}$ by their connecting Hamiltonian element 
$\bra{\mathrm{D}_{\mathbf{n}}}\hat{H}\ket{\mathrm{D}_{\mathbf{m}}}$ as part of 
the \textit{spawn} step and the algorithms 
to choose $\ket{\mathrm{D}_{\mathbf{n}}}$ given 
$\ket{\mathrm{D}_{\mathbf{m}}}$ are called excitation generators. 
The probability of this generation is denoted
$p(\mathbf{n}|\mathbf{m})$ = $p_{\mathrm{gen}}$ after which a spawn occurs with 
probability $p_\mathrm{spawn} \propto \delta \tau \frac{|\bra{\mathrm{D}_{\mathbf{n}}}\hat{H}\ket{\mathrm{D}_{\mathbf{m}}}|}{p_{\mathrm{gen}}}$,
with time step $\delta \tau$.
\par For an efficient calculation, $p_\mathrm{spawn}$ should have a reasonable value.
If $p_\mathrm{spawn}>1$, multiple particles are spawned at the 
same time, known as a ``bloom'', which is undesirable as it leads to 
less controllable population dynamics.
If, however, $p_\mathrm{spawn}$ is small, determinants are selected with no resulting spawn, and the algorithm is 
inefficient. $p_\mathrm{spawn}$ therefore ideally has a constant value,
which can be altered by the time step $\delta \tau$. Hence, it is desirable to make $p_{\mathrm{gen}}$ proportional to
$|\bra{\mathrm{D}_{\mathbf{n}}}\hat{H}\ket{\mathrm{D}_{\mathbf{m}}}|$ rather than
selecting determinants uniformly. Holmes et al. \cite{Holmes2016a} have 
introduced a heat bath sampling algorithm which weights the probability of 
choosing $\ket{\mathrm{D}_{\mathbf{n}}}$ with approximately
$\bra{\mathrm{D}_{\mathbf{n}}} \hat{H} \ket{\mathrm{D}_{\mathbf{m}}}$, but  
requires pre-computation of Hamiltonian elements leading to a significant 
storage cost of $\mathcal{O}(M^4)$ (which is of the same order as stored 
integrals if the code does not calculate them on-the-fly) and computational cost 
of $\mathcal{O}(N)$ where $M$ and $N$ are the size of the 
basis set and number of electrons respectively. Smart et al. \cite{Smartunpub} 
have reported the use of the Cauchy--Schwarz-like inequalities to 
provide upper bounds for $\bra{\mathrm{D}_{\mathbf{n}}} H \ket{\mathrm{D}_{\mathbf{m}}}$ with 
weights calculated on-the-fly. This reduces the storage cost while being 
linearly scaling in the number of orbitals.
\par Inspired by these ideas, excitation generators were investigated with weights generated
on-the-fly using Cauchy--Schwarz and
Power--Pitzer \cite{Power1974} inequalities to approximate 
$|\bra{\mathrm{D}_{\mathbf{n}}}\hat{H}\ket{\mathrm{D}_{\mathbf{m}}}|$ whose computational cost scales linearly with the number
of spinorbitals in the basis, $M$. We then present a new excitation generator that
uses this Power--Pitzer inequality but is of low computational order, 
$\mathcal{O}(N_\mathrm{ex.})$ in the case of CCMC or $\mathcal{O}(N)$ when using FCIQMC,
with memory cost $\mathcal{O}(M^2)$ which is also below the heat bath memory
scaling. $N_\mathrm{ex.}$ for a determinant or excitor is the number of electrons excited with respect to the reference .  For a truncated coupled cluster theory $N_\mathrm{ex.}$ does not scale with system size.
In a single-reference calculation, the reference determinant carries the most weight in the wavefunction and the majority of spawnings occur from determinants within a few electrons of excitation of this.
We therefore may pre-compute excitation weightings based on the reference
determinant, which shares the majority of electrons with nearby excited determinants, and then map the excitation to apply to any excited determinant, $\ket{\mathrm{D}_{\mathbf{n}}}$.
By this method, similar weights to the heat bath algorithm are employed and the
spread in 
$\frac{|\bra{\mathrm{D}_{\mathbf{n}}}\hat{H}\ket{\mathrm{D}_{\mathbf{m}}}|}{p_{\mathrm{gen}}}$ 
is minimised at a reduced computational and memory cost.
\par We now give a brief summary of the CCMC method, followed by a more detailed description
of various excitation generators whose performances we then compare.

\section{Coupled Cluster Monte Carlo}
This section describes the coupled cluster Monte Carlo (CCMC) method.
Full configuration interaction Quantum Monte Carlo (FCIQMC) has been discussed
extensively in the literature, see e.g. Refs. \cite{Booth2009,Cleland2010,Booth2013}.
CCMC solves the coupled cluster equations 
stochastically enabling calculations with larger basis sets and coupled 
cluster truncation levels than deterministic methods.
This section gives an overview over the algorithm and 
more information can be found in 
Refs. \cite{Thom2010,Spencer2016,Franklin2016,Scott2017b}.
\par The single reference coupled cluster wavefunction $\ket{\Psi}$ is written as
\begin{equation}
\ket{\Psi} \propto \exp{(\hat{T})} \ket{\mathrm{D}_{\mathbf{0}}}
\end{equation}
where $\ket{\mathrm{D}_{\mathbf{0}}}$ is the reference determinant and
\begin{equation}
\hat{T} = \sum_{\mathbf{i}}t_{\mathbf{i}}\hat{a}_{\mathbf{i}}.
\end{equation}
$\hat{a}_{\mathbf{i}}$ are ``excitors'' that generate determinants from the 
reference as
\begin{equation}
\ket{\mathrm{D}_{\mathbf{i}}} = 
    \hat{a}_{\mathbf{i}}\ket{\mathrm{D}_{\mathbf{0}}}.
\end{equation}
The one-electron orbitals used here are all orthogonal. The set of 
$\hat{a}_{\mathbf{i}}$ included depends on the coupled cluster truncation 
level. In coupled cluster singles and doubles (CCSD), those that 
create single and double excitations are included whereas CCSDT contains those with triple 
excitations as well and so on.
The unconventional unlinked coupled cluster equations are solved as
\cite{Helgaker2000},
\begin{equation}
\bra{\mathrm{D}_{\mathbf{n}}} \hat{H} \ket{\Psi} = E 
    \braket{\mathrm{D}_{\mathbf{n}}|\Psi}
\end{equation}
for the ground state energy $E$. Multiplying by a small number, $\delta \tau$, 
this can be rewritten as
\begin{equation}
\bra{\mathrm{D}_{\mathbf{n}}} 1 - \delta\tau(\hat{H}-E)\ket{\Psi} = 
    \braket{\mathrm{D}_{\mathbf{n}} | \Psi}.
\end{equation}
with imaginary time step $\delta\tau$. We arrive at an iterative equation for 
the coefficients $t_{\mathbf{i}}$ for the Slater determinants in $\Psi$ in 
imaginary time $\tau$
\begin{equation}
t_{\mathbf{n}}(\tau+\delta\tau) = t_{\mathbf{n}}(\tau) - \delta\tau 
    \bra{\mathrm{D}_{\mathbf{n}}}(\hat{H}-E)\ket{\Psi(\tau)}.
\label{eq:iterative}
\end{equation}
Franklin et al. \cite{Franklin2016} have rewritten equation \ref{eq:iterative} 
as
\begin{equation}
\begin{split}
t_{\mathbf{n}}(\tau+\delta\tau) = & t_{\mathbf{n}}(\tau) \\ &- \delta\tau 
    \bra{\mathrm{D}_{\mathbf{n}}}(\hat{H}-E_{\mathrm{proj.}}-E_{\mathrm{HF}})\ket{\Psi(\tau)} 
    \\&- \delta\tau(E_{\mathrm{proj.}} - S) t_{\mathbf{n}}(\tau).
\end{split}
\label{eq:iterative2}
\end{equation}
where the sum of the Hartree--Fock energy $E_{\mathrm{HF}}$ 
and the shift $S$ was substituted for the ground state energy $E$. The projected energy 
$E_{\mathrm{proj.}}$ and the population controlling shift $S$ (described below) 
are both measures for the correlation energy and are relatively uncorrelated. 
$E_{\mathrm{proj.}}$ is given by
\begin{equation}
E_{\mathrm{proj.}} = 
    \frac{\bra{\mathrm{D}_\mathbf{0}}\hat{H}-E_{\mathrm{HF}}\ket{\Psi}} 
    {\braket{\mathrm{D}_\mathbf{0}|\Psi}}.
\label{eq:proje}
\end{equation}
Equation \ref{eq:iterative2} is then sampled stochastically as described below 
and $t_{\pmb{\mathrm{n}}}$ is updated at each time step. Monte Carlo 
particles, ``excips'', are placed on the excitors $a_{\pmb{\mathrm{i}}}$. 
At first, all excips are on the null excitor $a_{\pmb{\mathrm{0}}}$ that gives 
back the reference determinant. As the simulation proceeds, they multiply and spread to other excitors 
with \textit{spawn}, \textit{death/birth} and \textit{annihilation} steps 
at each imaginary time step.
\par During each time step, a single excitor
or cluster of excitors which have particles on them are first randomly selected, e.g. the two excitors
$a_{\pmb{\mathrm{i}}}$ and $a_{\pmb{\mathrm{j}}}$, that when acting together on the reference 
determinant $\ket{\mathrm{D}_{\pmb{\mathrm{0}}}}$, gives another determinant 
$\ket{\mathrm{D}_{\pmb{\mathrm{m}}}}$, i.e. $\hat{a}_{\pmb{\mathrm{i}}} 
\hat{a}_{\pmb{\mathrm{j}}}\ket{\mathrm{D}_{\pmb{\mathrm{0}}}} = \hat{a}_{\pmb{\mathrm{i}}} 
\ket{\mathrm{D}_{\pmb{\mathrm{j}}}} = \ket{\mathrm{D}_{\pmb{\mathrm{m}}}}$.
This determinant then undergoes the following processes:
\begin{itemize}
        \item \textit{Spawn:} Another determinant 
            $\ket{\mathrm{D}_{\pmb{\mathrm{n}}}}$ is randomly selected with a probability 
            $p_{\mathrm{gen}}$. An excip of appropriate sign is placed on 
            $\hat{a}_{\mathbf{n}}$ with a probability proportional to 
            $\frac{\delta\tau|\bra{\mathrm{D}_{\pmb{\mathrm{n}}}} \hat{H} 
            \ket{\mathrm{D}_{\pmb{\mathrm{m}}}}|}{p_{\mathrm{gen}}}$.
        \item \textit{Death/Birth:} An excip of opposite or the same sign is 
            placed on $a_{\mathbf{m}}$ with a probability proportional to 
            $|\bra{\mathrm{D}_{\mathbf{m}}} \hat{H} - S - E_{\mathrm{HF}} 
            \ket{\mathrm{D}_{\mathbf{m}}}|$ if just one excitor was used to form 
            $\ket{\mathrm{D}_{\pmb{\mathrm{m}}}}$ and a probability 
            proportional to $|\bra{\mathrm{D}_{\mathbf{m}}} \hat{H} - 
            E_{\mathrm{proj.}} - E_{\mathrm{HF}} \ket{\mathrm{D}_{\mathbf{m}}}|$ 
            if a cluster was used.
        \item \textit{Annihilation:} Finally, at the end of a imaginary time 
            step, excip pairs of opposite sign on the same excitor are removed.
\end{itemize}
The shift is initially set to zero and is allowed to vary once a the total 
population (number of particles), $N_\mathrm{ex.}$, is higher than the critical population at the ``shoulder'' or 
``plateau'' \cite{Thom2010,Spencer2012}. To give an on-average constant population, it is updated every $B$ iterations 
according to
\begin{equation}
S(\tau) = S(\tau - \delta\tau B) - 
    \frac{\gamma}{B\delta\tau}\ln{\frac{N(\tau)}{N(\tau- \delta\tau B)}}
\end{equation}
where $\gamma$ is the shift damping factor.
\par Rather than integer-valued, real-valued excip amplitudes \cite{Petruzielo2012,Overy2014} have been used and
the full non-composite version of the CCMC algorithm \cite{Spencer2018} with 
truncated and even selection \cite{Scott2017b} has been applied. We have also used parallelization as described in
Ref. \cite{Spencer2018}. The results here were checked 
for population control bias using a reweighting scheme by Umrigar et al. 
\cite{Umrigar1993} and Vigor et al. \cite{Vigor2015}. Data has been reblocked \cite{Flyvbjerg1989}
implemented in pyblock \footnote{For code, 
see \url{https://github.com/jsspencer/pyblock}} to estimate error bars. Our CCMC and FCIQMC calculations were done 
with the HANDE code \cite{HANDEpaper} which is open source\footnote{See 
\url{http://www.hande.org.uk/} and \url{https://github.com/hande-qmc/hande} for 
information and code}.

\section{Excitation Generators}
As mentioned above, in the \textit{spawn} step, the excitation generator selects a determinant 
$\ket{\mathrm{D}_{\pmb{\mathrm{n}}}}$ connected to 
$\ket{\mathrm{D}_{\mathbf{m}}}$ with probability 
$p_{\mathrm{gen}}$. The spawn probability is proportional to 
$\frac{\delta\tau|\bra{\mathrm{D}_{\pmb{\mathrm{n}}}} \hat{H} 
\ket{\mathrm{D}_{\pmb{\mathrm{m}}}}|}{p_{\mathrm{gen}}}$.
In this paper, we present a 
method that aims to use an optimal $p_{\mathrm{gen}}$ so that more important 
determinants are selected with a higher probability. An introduction to 
excitation generators in FCIQMC which uses the same/similar excitation 
generators, is given by Booth et al. \cite{Booth2009,Booth2014}. The idea of 
excitation generation and dividing by the generation probability was also 
discussed in e.g. Refs. \cite{Holmes2016a,Thom2005,Thom2007,Kolodrubetz2012, Pederiva2017} and a transition 
with uniform selection is also done by the configuration state function 
projector Monte Carlo method of Ohtsuka et al. \cite{Ohtsuka2008}. Kolodrubetz et al. \cite{Kolodrubetz2012}
used a weighted excitation generator that --- among other distributions --- used the inverse momentum squared as a weight.
Booth et al. \cite{Booth2014} also considered weighting the excitation generation by Hamiltonian matrix elements by
enumerating a subset of excitations with the magnitudes of these Hamiltonian elements. Due to the cost of finding
$p_\mathrm{gen.}$, this idea was not pursued further.
 A version of 
the uniform excitation generators described here, is explained in detail in Ref. 
\cite{Booth2014}.
\par The spawn probability is only non-zero if $\bra{\mathrm{D}_{\mathbf{n}}} 
\hat{H} \ket{\mathrm{D}_{\mathbf{m}}}$ is non-zero. The Hamiltonians, $\hat{H}$, 
 considered here only contain constant, one body, and two body terms.  
$\bra{\mathrm{D}_{\mathbf{n}}} \hat{H} \ket{\mathrm{D}_{\mathbf{m}}}$ can therefore only 
be non-zero if $\ket{\mathrm{D}_{\mathbf{n}}}$ and 
$\ket{\mathrm{D}_{\mathbf{m}}}$ differ by at most two orbitals. To 
select a suitable $\ket{\mathrm{D}_{\pmb{\mathrm{n}}}}$ for 
$\ket{\mathrm{D}_{\pmb{\mathrm{m}}}}$ to spawn to, we can create a single or a 
double excitation from $\ket{\mathrm{D}_{\pmb{\mathrm{m}}}}$ to generate 
$\ket{\mathrm{D}_{\pmb{\mathrm{n}}}}$ ($\pmb{\mathrm{n}} \neq 
\pmb{\mathrm{m}}$). Any other excitation would lead to a zero spawn probability. 
Except for the ``original'' heat bath excitation generator, all excitation 
generators discussed here create a single or double excitation from 
$\ket{\mathrm{D}_{\pmb{\mathrm{m}}}}$ to generate 
$\ket{\mathrm{D}_{\pmb{\mathrm{n}}}}$ with probability $p_\mathrm{single}$ or 
$1-p_\mathrm{single}$ respectively. As suggested by Holmes et al. \cite{Holmes2016a},
we aim to appropriately select $p_{\mathrm{spawn}, \mathrm{single}}$ and
$p_{\mathrm{spawn}, \mathrm{double}}$ by setting $p_\mathrm{single}$ suitably to optmize the distribution of excitations.
For a single excitation 
where electron in spinorbital $i$ is excited to spinorbital $a$,
\begin{equation}
p_{\mathrm{gen, single}} = p_\mathrm{single}p_\mathrm{method}p(i)p(a|i).
\end{equation}
where $p_\mathrm{method}$ contains additional factors depending 
on the selection method of $i$ and $a$. 
\par In the case of a double 
excitation, $ij \rightarrow ab$, as $i$ and $j$ ideally come from the same set 
of orbitals (those occupied in the determinant) and so do $a$ and $b$ (those unoccupied in the determinant), 
first $ij$ and then $ab$ are selected in all excitation generators discussed 
here. That means that for example while the selection order between $i$ and $j$ 
can vary, $a$ will not be selected before either $i$ and $j$. The possible 
orders are therefore $ijab$, $ijba$, $jiab$ and $jiba$. While the first selected occupied is called
$i$ and the second $j$, their indistinguishability  has to be taken into account 
when calculating $p_{\mathrm{gen}}$:
\begin{equation}
\begin{split}
p_{\mathrm{gen, double}} = \\(1-p_\mathrm{single})p_\mathrm{method} 
    (p(i)p(j|i)p(a|i,j)p(b|a,i,j) + \\ p(i)p(j|i)p(b|i,j)p(a|b,i,j) + \\ 
    p(j)p(i|j)p(a|j,i)p(b|a,j,i) + \\ p(j)p(i|j)p(b|j,i)p(a|b,j,i)).
\end{split}
\end{equation}
\par In a rather basic implementation, the spinorbitals with electrons 
to excite $i$ (and $j$) and the spinorbitals to excite to $a$ (and $b$) are selected with 
uniform probabilities. The excitation generator that we call \textit{not 
renormalised excitation generator} or simply \textit{no. renorm.} here, when 
doing a double excitation, first selects $i$ and $j$ as a pair with uniform 
probability from the set of occupied orbitals. In that case,
\begin{equation}
        p_\mathrm{method}(p(i)p(j|i) + p(j)p(i|j)) = \frac{2}{N(N-1)},
\end{equation}
where the number of electron is $N$. If both $i$ and $j$ have the same spin, $\sigma$, then $a$ 
is uniformly chosen from the set of virtual orbitals of that spin, otherwise it can be any virtual orbital. 
$b$ is then selected uniformly from the set of orbitals (excluding $a$) with required spin and 
symmetry. With this 
selection of $b$, it is possible that after the selection of $i$, $j$, and $a$, 
there are no possible selections of $b$, it is a forbidden excitation generation. In 
that case the spawn attempt will be unsuccessful (we set 
$\bra{\mathrm{D}_{\mathbf{m}}} H \ket{\mathrm{D}_{\mathbf{n}}}$ = 0).
\par The
choice of how to select which electrons to excite and to which spinorbitals they are excited is 
is entirely arbitrary (assuming all valid excitations are possible), as long as the probability with which this 
selection has been done is known and $p_{\mathrm{gen}}$ is then calculated accordingly.
As an alternative to the \textit{not renormalised excitation generator} 
(\textit{no. renorm.}), forbidden 
excitations (which lead to unsuccessful spawns) can be avoided by generating a different excitation and
renormalising the appropriate probabilities. This is called the 
\textit{renormalised excitation generator} or in short, \textit{renorm.}. Again, 
see Booth et al. \cite{Booth2009,Booth2014} for an in-depth description of uniform 
excitation generators.
\par In the following subsections, we describe the heat bath excitation
generators and the 
\textit{heat bath/uniform Power--Pitzer} excitation generators which follow the 
ideas of Alavi and others. Finally, the 
\textit{heat bath Power--Pitzer ref.} excitation generator is presented, which pre-computes 
some weights based on the reference determinant and therefore has a very low
computational cost not scaling with system size ($\mathcal{O}(N_\mathrm{ex.})$ when using CCMC or
scaling as $\mathcal{O}(N)$ for FCIQMC instead of $\mathcal{O}(M)$). Its memory 
cost is significantly less than \textit{heat bath} excitation generators, being 
 $\mathcal{O}(M^2)$ instead of $\mathcal{O}(M^4)$. In appendix \ref{app:furtheruni}, further uniform excitation
generators are discussed.
\par Table \ref{tab:weighted} gives an overview over the weighted excitation generators presented here.
This table should be understood together with the following descriptions in the next subsections.
\begin{table*}
	\caption{Overview of weighted excitation generators. C.--S. means Cauchy--Schwarz and P.--P. Power--Pitzer.
	p.c. is pre-calculated and o.t.f. means on-the-fly. Comp./memory $\mathcal{O}$ is the computational/memory order
	the excitation generator scales with.
	As a method of selection, ``heat bath'' refers to ``selecting those like the heat bath excitation generator''.
	Single excitations or $ij$ in a double excitation that have been selected ``uniformly'', have been selected
	with the uniform \textit{renorm.} excitation generator. $N$ is number of electrons, $M$ the number of spinorbitals
	and $N_\mathrm{ex.}$ the excitation level possible from the reference at this coupled cluster level.}
		\begin{tabular}{|l|lllll|}
		\hline
			& single excitations & $ij$ & $ab$ & comp. $\mathcal{O}$ & memory $\mathcal{O}$  \\
			\hline
			\textit{heat bath} & decision after & heat bath & heat bath & $N$ & $M^4$  \\
			 & having selected \textit{ija} &  &  &  &   \\
			\textit{heat bath uniform singles} & uniformly & heat bath & heat bath & $N$ & $M^4$  \\
			\textit{heat bath exact singles} & exactly, on-the-fly & heat bath  & heat bath & $NM$ & $M^4$  \\
			\textit{uniform Cauchy--Schwarz} & uniformly & uniformly & C.--S. o.t.f. &$M$ & $M$  \\
			\textit{uniform Power--Pitzer} & uniformly & uniformly & P.--P. o.t.f. &$M$ & $M$  \\
			\textit{heat bath Cauchy--Schwarz} & uniformly & heat bath  & C.--S. o.t.f.&$M$ & $M^2$  \\
			\textit{heat bath Power--Pitzer} & uniformly & heat bath  & P.--P. o.t.f.&$M$ & $M^2$  \\
			\textit{heat bath Power--Pitzer ref.} & p.c. & heat bath p.c. & P.--P. p.c. &$N$ or $N_\mathrm{ex.}$ & $M^2$ \\
			\hline
		\end{tabular}
	\label{tab:weighted}
\end{table*}

\subsection{Heat Bath Excitation Generators}
The heat bath excitation generators aim to get the orbital selection weights as 
close as possible to the Hamiltonian matrix element 
$|\bra{\mathrm{D}_{\mathbf{n}}} \hat{H} \ket{\mathrm{D}_{\mathbf{m}}}|$ with the 
aim of making part of the spawn probability 
$\frac{|\bra{\mathrm{D}_{\mathbf{n}}} \hat{H} 
\ket{\mathrm{D}_{\mathbf{m}}}|}{p_\mathrm{gen}}$ as close as possible to a 
constant. In the case of a double excitation $ij \rightarrow ab$,
$p_\mathrm{gen}$ can be rewritten as
\begin{equation}
\begin{split}
p_\mathrm{gen, double} = \\p(i)\times p(j|i) \times p(a|ij) \times p(b|ija) = \\ 
    \frac{\sum_{jab}H_{ijab}}{\sum_{ijab}  H_{ijab}} \times 
    \frac{\sum_{ab}H_{ijab}}{\sum_{jab} H_{ijab}} \times 
    \frac{\sum_{b}H_{ijab}}{\sum_{ab} H_{ijab}} \times \frac{H_{ijab}}{\sum_{b} 
    H_{ijab}},
\end{split}
\label{eq:hb_basic}
\end{equation}
where $H_{ijab} = |\bra{\mathrm{D}_{\mathbf{n}}} \hat{H} 
\ket{\mathrm{D}_{\mathbf{m}}}|$ where $\ket{\mathrm{D}_{\mathbf{m}}}$ and 
$\ket{\mathrm{D}_{\mathbf{n}}}$ differ by the excitation $ij \rightarrow ab$. In 
the heat bath excitation generators,
$\frac{\sum_{jab}H_{ijab}}{\sum_{ijab} H_{ijab}}$ is an approximation for $p(i)$ and so on.
\par Here, we distinguish between three different heat bath excitation 
generators described by/based on Holmes et al. \cite{Holmes2016a}. The 
``original'' heat bath excitation generator as introduced and described in detail 
by Holmes et al. \cite{Holmes2016a} (in short \textit{heat bath}), the heat bath 
excitation generator that decides first whether a single or a double 
excitation is performed and which samples singles uniformly which is mentioned by Holmes et 
al. \cite{Holmes2016a} (\textit{heat bath uniform singles}) and finally, the one that 
first decides whether to do a single or double excitation and samples singles 
exactly according to their Hamiltonian matrix element, \textit{heat bath 
exact singles}\footnote{Idea by Alavi and co--workers, this was suggested to us as an alternative by Pablo L\'opez R\'ios
(personal communication).}. For more information and an in-depth description, see Ref.
\cite{Holmes2016a}.
\par In all three heat bath excitation generators, 
all possible contractions of $H_{ijab}$ appearing in equation \ref{eq:hb_basic} are
pre-computed and stored. 
More specifically, $H_i = \sum_{jab} H_{ijab}$, $H_{ij} = \sum_{ab} H_{ijab}$, 
$H_{ija} = \sum_{b} H_{ijab}$ and $H_{ijab}$ are pre-computed where $i, j, a$ and $b$ can be any 
spinorbital in the calculation. In all sums $i \neq j \neq a \neq b$. The 
alias methods\cite{Walker1974,Walker1977,Kronmal1979,Holmes2016a} are used and alias tables are pre-calculated for selecting $a$ 
(given $ij$) with weights $H_{ija}$ and one for selecting $b$ (given $ija$) with 
weights $H_{ijab}$ (which is of $\mathcal{O}(M^4)$). The look-up time with the alias 
method is of $\mathcal{O}(1)$. The alias tables for selecting $i$ and selecting 
$i$ given $j$ are computed on-the-fly using pre-computed weights in 
$\mathcal{O}(N)$ time. The alias table for selecting $i$ then only considers 
$H_i$ from the set of occupied orbitals for $i$ and when selecting $j$ given 
$i$, the alias table only considers $H_{ij}$ with occupied $j$.
\par When using the \textit{heat bath} excitation generator to find an 
excitation, first an alias table is created on-the-fly for $i$ as described above 
and then $i$ is selected. We proceed similarly for $j$. Using the pre-computed alias 
table with weights $H_{ija}$, $a$ is found. If this orbital is occupied, we have 
a forbidden excitation and the spawn attempt was unsuccessful. Only at this 
stage it is decided whether to attempt a single or a double excitation. In the 
algorithm by Holmes et al. \cite{Holmes2016a}, a single excitation is attempted 
with probability $\frac{H_{ia}}{H_{ia} + H_{ija}}$ and a double excitation is 
attempted with probability $\frac{H_{ija}}{H_{ia} + H_{ija}}$ if $H_{ia} < 
H_{ija}$ where $H_{ia} = |\bra{\mathrm{D}_{\mathbf{m}}} \hat{H} 
\ket{\mathrm{D}_{\mathbf{k}}}|$ with $\ket{\mathrm{D}_{\mathbf{m}}}$ and 
$\ket{\mathrm{D}_{\mathbf{k}}}$ connected by the excitation $i \rightarrow a$. 
However, if $H_{ia} > H_{ija}$, both a double and a single excitation are 
attempted \footnote{It is not clear from Holmes et al. \cite{Holmes2016a} what 
happens if $H_{ia} = H_{ija}$}. This avoids low probabilities for choosing to do 
a double excitation if $H_{ia}$ gets large. In our implementation  
 in HANDE \cite{HANDEpaper}, that approach was modified to 
only allow one excitation attempt per excitation generator call. If $H_{ia} \geq 
H_{ija}$, instead of choosing to attempt a single ($i \rightarrow a$) and a 
double ($ij \rightarrow ab$) excitation, a single or a double 
excitation is attempted with probability $\frac{1}{2}$ respectively. The rest follows Holmes 
et al. \cite{Holmes2016a}. Either a single excitation $i 
\rightarrow a$ is attempted now or $b$ is selected from pre-computed weights and a double 
excitation $ij \rightarrow ab$ (provided $b$ is not occupied) is attempted.
\par The \textit{heat bath} excitation generator relies on single excitations 
being less significant. It has the major drawback in that is potentially has a 
bias if there exists no $j$ to be selected after $i$ and before $a$ if $i 
\rightarrow a$ is valid. This is explained in more detail in Ref. 
\cite{Holmes2016a}. Our conservative but robust test for bias as implemented in
HANDE, counts the number of $j$ for which $\sum_{b}H_{ijab}$ is non zero for given
$ia$. If this number is greater than the number of virtual orbitals,
then there will always be an occupied $j$ to be selected for allowed $i \rightarrow a$ and
there is no bias.

\subsection{On-the-fly Power--Pitzer Excitation Generators}
While bringing $\frac{|\bra{\mathrm{D}_{\mathbf{n}}} \hat{H} 
\ket{\mathrm{D}_{\mathbf{m}}}|}{p_\mathrm{gen}}$ closer to a constant as uniform 
excitation generators \cite{Holmes2016a}, \textit{heat bath} 
excitation generators suffer from a large memory cost ($\mathcal{O}(M^4)$). To 
reduce the memory cost, Alavi and Smart et al. \cite{Smartunpub} had the idea of 
calculating approximate weights on-the-fly in $\mathcal{O}(M)$ calculation time. 
This is for example mentioned by Blunt et al. \cite{Blunt2017} and Holmes et al. 
\cite{Holmes2016a}. They proposed calculating Cauchy--Schwarz-like
 upper bounds on the two body part of the Hamiltonian on-the-fly 
when doing a double excitation. Here, we also describe an excitation generator that uses
an inequality derived by Power and Pitzer\cite{Power1974} instead. It effectively differs 
from \textit{Cauchy--Schwarz} excitation generators by the usage of exchange rather than 
Coulomb integrals. We note that the \textit{Cauchy--Schwarz} excitation generators
mentioned here may not quite replicate excitation generators of Alavi et al.
\footnote{Personal communication with Ali Alavi and Pablo L\'opez R\'ios.} which are yet to be fully published.
\par Given that $i$, $j$, $a$ and $b$ are different, the only non-zero part of 
the Hamiltonian element $\bra{\mathrm{D}_{\pmb{\mathrm{m}}}} H 
\ket{\mathrm{D}_{\pmb{\mathrm{n}}}}$ in a double excitation are the Coulomb 
integral $\braket{ij|ab}$ and the exchange integral $\braket{ij|ba}$ according 
to Slater-Condon rules \cite{Slater1929,Condon1930}.
Here, the notation
\begin{equation}
\braket{ij|ab}=\int 
    \frac{\phi^*_i(\pmb{r}_1)\phi^*_j(\pmb{r}_2)\phi_a(\pmb{r}_1)\phi_b(\pmb{r}_2) 
    \mathrm{d}\pmb{r}_1\mathrm{d}\pmb{r}_2}{|\pmb{r}_1-\pmb{r}_2|},
\end{equation}
is used with one electron orbitals/spinorbitals $\phi$ that make up Slater determinants 
$\ket{D_\mathbf{x}}$.
en example of such a weight used by Alavi and others for $ij \rightarrow ab$ is 
a Cauchy--Schwarz upper bound on $\braket{ij|ab}$ given by 
\begin{equation}
\sqrt{|\braket{ia|ia}|\ |\braket{jb|jb}|} \geq |\braket{ij|ab}|.
\end{equation}
\par The weights are such that $a$ can be chosen
(almost) independently of $b$ and vice versa which makes the algorithm linear 
scaling in the number of spinorbitals. A Power--Pitzer \cite{Power1974} inequality (derived previously
for real wavefunctions\cite{Whitten1973}) is
\begin{equation}
\sqrt{|\braket{ia|ai}|\ |\braket{jb|bj}|} \geq |\braket{ij|ab}|.
\end{equation}
Exchange integrals are lower or equal in magnitude than Coulomb 
integrals (see e.g. Ref. \cite{Roothaan1951}) which means that exchange integrals are the tighter upper 
bound for $|\braket{ij|ab}|$. The two body 
term in the Hamiltonian is $\braket{ij|ab} - \braket{ij|ba}$. When $a$ and $b$ 
have opposite spin, the two body term reduces to $\braket{ij|ab}$ and its 
Power--Pitzer upper bound is used as the weight. If $a$ and $b$ have the same 
spin, both orderings, $ab$ and $ba$ will generate the same excitation, and this is included in $p_\mathrm{gen}$.
This section gives a detailed description of the algorithm.
\par $i$ and $j$ can be selected uniformly or with the \textit{heat bath}
weightings producing a family of excitation generators.
We denote by \textit{uniform Cauchy--Schwarz} and \textit{uniform Power--Pitzer}
excitation generators which select them uniformly, like 
the \textit{renorm.} excitation generator, and by 
\textit{heat bath Cauchy--Schwarz} and \textit{heat bath Power--Pitzer} those which select
them as the \textit{heat bath} excitation 
generators do with pre-calculated weights with memory cost of $\mathcal{O}(M^2)$\footnote{The idea of selecting $ij$
like the \textit{heat bath} excitation generator was communicated by Pablo L\'opez R\'ios
(personal communication).}. The computational scaling is $\mathcal{O}(M)$ in both cases.
\par The  \textit{Power--Pitzer}  and \textit{Cauchy--Schwarz} excitation generators first decide whether to attempt a single or a double 
excitation according to  $p_\mathrm{single}$. For single excitations, the
\textit{renorm.} excitation generator is employed. When attempting double 
excitations, $i$ and $j$ are selected either uniformly or with \textit{heat bath} 
weights out of the occupied orbitals of $\ket{\mathrm{D}_{\pmb{\mathrm{m}}}}$. 
Then, $a$ is selected out of the set of virtual spinorbitals 
$a_{\sigma_i,\mathrm{virt.}}$ with the same spin as $i$. $a$ is selected with the 
probability of
\begin{equation}
p(a|ij)=p(a|i) = 
    \frac{\sqrt{|\braket{ia|ai}|}}{\sum_{a=a_{\sigma_i,\mathrm{virt.}}} 
    \sqrt{|\braket{ia|ai}|}}
\end{equation} 
when using \textit{Power--Pitzer} excitation generators or
\begin{equation}
p(a|ij)=p(a|i) = 
    \frac{\sqrt{|\braket{ia|ia}|}}{\sum_{a=a_{\sigma_i,\mathrm{virt.}}} 
    \sqrt{|\braket{ia|ia}|}}
\end{equation} 
when using \textit{Cauchy--Schwarz} excitation generators.
$b$, the second orbital to excite to, it selected out of the set 
of spinorbitals $b_{\neq a, \sigma_j,\mathrm{sym.}}$ of the same 
spin as $j$ and the required symmetry to conserve overall symmetry and not equal 
to $a$.
The weights are given by $\braket{jb|jb}$ (\textit{Cauchy--Schwarz}) or
$\braket{jb|bj}$ (\textit{Power--Pitzer}). If the total weight when 
finding $b$ is zero (i.e. there are no spinorbitals with the required spin and 
symmetry or only the spinorbitals found as $a$ has that spin and symmetry) or if 
the found $b$ is already occupied, the spawn attempt is unsuccessful.
Again, orbitals $a$ and $b$ were selected using their weights with the alias 
method\cite{Walker1974,Walker1977,Kronmal1979,Holmes2016a}.
\par The performance of the four excitation generators described in this
subsection, \textit{uniform Cauchy--Schwarz}, \textit{heat bath Cauchy--Schwarz},
\textit{uniform Power--Pitzer}, and \textit{heat bath Power--Pitzer}, were then tested, using a chain of three water molecules in the cc-pVDZ basis
\cite{Dunning1989}, whose molecular orbitals have been localized.
The excitation generators all come with a low memory cost, 
which is $\mathcal{O}(M)$ temporarily or $\mathcal{O}(M^2)$ and all scales
as $\mathcal{O}(M)$ in computational time. 
%Since they have the same computational
%and different but relatively low memory scaling, it is compared how close
The distribution of $\frac{|\bra{\mathrm{D}_{\mathbf{n}}} \hat{H} \ket{\mathrm{D}_{\mathbf{m}}}|}{p_\mathrm{gen}}$, which should ideally be constant, was compared for the four excitation generators.
\begin{figure}
\centering
        \begin{subfigure}[h]{8.5cm}
        \includegraphics[width=1.0\linewidth,keepaspectratio]{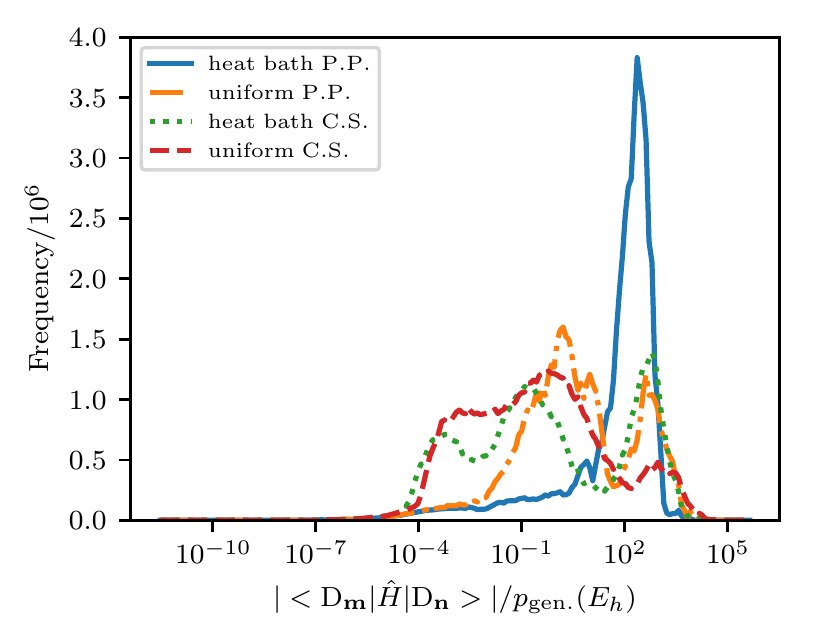}
        \end{subfigure}
        \newline
        \begin{subfigure}[h]{8.5cm}
        \includegraphics[width=1.0\linewidth,keepaspectratio]{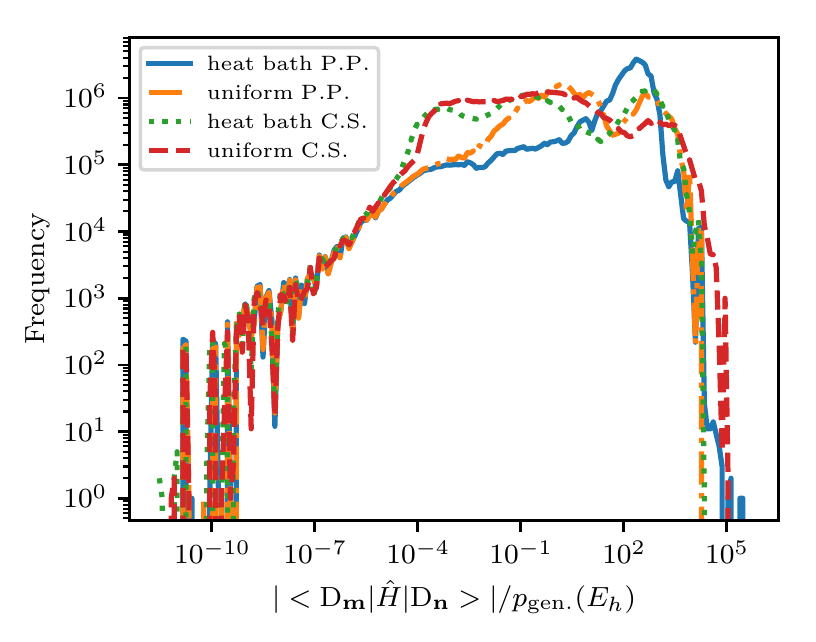}
        \end{subfigure}
        \newline
        \caption{Comparison of the histograms of $\frac{|\bra{\mathrm{D}_{\mathbf{n}}} \hat{H} \ket{\mathrm{D}_{\mathbf{m}}}|}{p_\mathrm{gen}}$ for the \textit{Cauchy--Schwarz} (C.S.)
        and
        \textit{Power--Pitzer} (P.P.) on-the-fly excitation generators. $ij$ are either selected uniformly or
        using heat bath. The computational scaling of all excitation generators here is $\mathcal{O}(M)$.
        CCSD was performed on three water molecules in the cc-pVDZ basis using localized MOs.
        The values were logged for one Monte Carlo iteration. The size of the bins is logarithmic.
        \textit{Bottom} graph took the
        log of the frequency whereas the \textit{top} graph did not. They both show the same data.
        All of them were restarted from the same calculation and then equilibrated before taking data.
        $\frac{|\bra{\mathrm{D}_{\mathbf{n}}} \hat{H} \ket{\mathrm{D}_{\mathbf{m}}}|}{p_\mathrm{gen}}=0$
        data is not shown which includes forbidden excitations.
        $p_\mathrm{single}$ was set to be the same when running which was corrected in post-processing to make
        the mean of finite $\frac{|\bra{\mathrm{D}_{\mathbf{n}}} \hat{H} \ket{\mathrm{D}_{\mathbf{m}}}|}{p_\mathrm{gen}}$
         for single and double excitations coincide which did not change $p_\mathrm{single}$ values by more than 30\%.}
        \label{fig:ppcshisto}
\end{figure}
Figure \ref{fig:ppcshisto} shows the histograms (excluding $\frac{|\bra{\mathrm{D}_{\mathbf{n}}} \hat{H} \ket{\mathrm{D}_{\mathbf{m}}}|}{p_\mathrm{gen}}=0$) with linear and logarithmic frequency scales.
The \textit{bottom} graph shows the all excitation
generators have similar looking tails to both sides, the \textit{heat bath Power--Pitzer}
having the longest tail to big $\frac{|\bra{\mathrm{D}_{\mathbf{n}}} \hat{H} \ket {\mathrm{D}_{\mathbf{m}}}|} {p_\mathrm{gen}}$. However, the number of events in bins above the maximum
$\frac{|\bra{\mathrm{D}_{\mathbf{n}}} \hat{H} \ket {\mathrm{D}_{\mathbf{m}}}|} {p_\mathrm{gen}}$ filled bin for the \textit{uniform Power--Pitzer} excitation generator --- which has the lowest maximum --- is
fewer than 100 events which is not significant relatively speaking so if not using initiator approximations
there should not be a noticeable effect. The \textit{top} graph demonstrates that the \textit{heat bath Power--Pitzer} 
gives the sharpest peak and makes $\frac{|\bra{\mathrm{D}_{\mathbf{n}}} \hat{H} \ket {\mathrm{D}_{\mathbf{m}}}|} {p_\mathrm{gen}}$  closest to a constant of the excitation generators.
\begin{table}
	\caption{Fraction of allowed and fraction of non-zero allowed spawn events, both with respect to total number
	of spawn events. The latter represents the spawn events depicted in figure \ref{fig:ppcshisto}. \textit{heat bath Cauchy--Schwarz} and \textit{uniform Cauchy--Schwarz} had similar values and the same for the \textit{Power--Pitzer}
	excitation generators, so they have been combined to \textit{C.--S.} and \textit{P.--P.} respectively.}

		\begin{tabular}{|l|l|l|}
		\hline
			& $\frac{\#\mathrm{allowed}}{\#\mathrm{total}}$ events & $\frac{\mathrm{\#allowed}\ \mathrm{non-zero}}{\mathrm{\#total}}$ events \\
			\hline
			\textit{C.--S.} & 0.8 & 0.69--0.70  \\
			\textit{P.--P.} & 0.68--0.69 & 0.68--0.69   \\ \hline
		\end{tabular}

	\label{tab:zerodisallowed}
\end{table} 
Only non-zero allowed events are shown in figure \ref{fig:ppcshisto}. Table \ref{tab:zerodisallowed} shows what
fraction that is of the total number of events (second column) and what fraction of events are allowed which includes the
 allowed but zero $\frac{|\bra{\mathrm{D}_{\mathbf{n}}} \hat{H} \ket {\mathrm{D}_{\mathbf{m}}}|} {p_\mathrm{gen}}$ events
 (first column). Both the \textit{Cauchy--Schwarz} and the \textit{Power--Pitzer} excitation generators have a similar
fraction of non-zero allowed events. The \textit{Power--Pitzer} excitation generators have more forbidden events but of
those that are allowed, more are non-zero. A big source for forbidden events is the selection of $b$ which is selected from
the set of orbitals of required spin and symmetry which can be occupied. An event is then forbidden if $b$ selected is
occupied. Our implementation could be further improved by excluding occupied orbitals from that selection.
In the results section we will let
\textit{heat bath Power--Pitzer} represent all these four excitation generators introduced in this
subsection.

\subsection{Pre-computed Power--Pitzer Excitation Generator}
Even with their reduced memory requirements, the above excitation generators still
add a considerable cost to calculations, and we seek a way to reduce this further. We now introduce an
$\mathcal{O}(N)$ \textit{Power--Pitzer} excitation generator, \textit{heat bath Power--Pitzer ref.} ,
where $N$ is the number of electrons. This can even be modified to be 
$\mathcal{O}(N_\mathrm{ex.})$ where $N_\mathrm{ex.}$ is the number of electrons excited
with respect to the reference if excitations instead of determinants were stored in our implementation.
Within a routine coupled cluster calculation, the maximum $N_\mathrm{ex.}$ does not depend on system size.
This excitation generator combines advantages of 
\textit{heat bath Power--Pitzer} where a bias check is not required
beforehand (but is with the ``original''\textit{heat bath} excitation generator) and
which has a significantly lower memory cost with the lower 
computational scaling of the \textit{heat bath} excitation 
generators, further improving upon this.
We make use of the single-reference nature of coupled cluster where 
the reference determinant $\ket{\mathrm{D}_{\mathbf{0}}}$ is more important than 
any other determinant by pre-computing some
weights based on the reference determinant. Pre-computed weights include heat bath and \textit{Power--Pitzer} weights,
for selecting the occupied and virtual orbitals respectively in a double excitation.
Spinorbitals are first found by pretending the reference determinant is the determinant we are
exciting from and are then mapped between the current determinant and 
the reference determinant when it is appropriate. The memory cost is 
$\mathcal{O}(M^2)$ while the computational cost when spawning is only the 
mapping of the reference $\ket{\mathrm{D}_{\mathbf{0}}}$ to the actual 
determinant $\ket{\mathrm{D}_{\mathbf{m}}}$ which is $\mathcal{O}(N)$.
Since weights are based on one determinant, it is not costly to
pre-calculate weights for single excitations as well. This is a considerable
advantage over the on-the-fly \textit{Power--Pitzer} and
\textit{heat bath} excitation generators that either do single excitations uniformly,
exactly (which is costly) or based on double excitation weights.
Note that while this section talks about single-reference systems,
this excitation generator also easily applies to systems that are multi-reference but
the Hamiltonian elements connecting the most important determinants to the other determinants
are similar.
\par In this algorithm, two frames of reference are considered. In the first frame, 
the reference frame, which is denoted by a prime, excitations are from the reference 
determinant, i.e. $\ket{\mathrm{D}_{\pmb{\mathrm{m'}}}} = 
\ket{\mathrm{D}_{\pmb{\mathrm{0}}}}$. In this frame, a double 
excitation would be $i'j' \rightarrow a'b'$. In the second frame, the simulation frame, 
the actual frame the calculation is in, excitations are from
$\ket{\mathrm{D}_{\pmb{\mathrm{m}}}}$ and that excitation is $ij \rightarrow ab$. 
For selecting some orbitals, the weights of the orbitals in the reference 
frame are used and its spinorbitals are mapped to the simulation frame to find the actual 
excitation as explained in appendix \ref{app:mapping}.
\par The following quantities for single 
excitations are pre-computed:
\begin{equation}
w_{i',\mathrm{s}} = \sum_a \left(\frac{1}{n_{jb}} \sum_{j=j_\mathrm{occ.ref.} 
    b=b_\mathrm{virt. ref.}} \bra{\mathrm{D}_{j}^{b}} \hat{H} 
    \ket{\mathrm{D}_{i'j}^{ab}}\right),
\end{equation}
where $i'$ is an occupied orbital in the reference and the sum over $a$ is over 
all orbitals with allowed excitation $i' \rightarrow a$. $n_{jb}$ is $N(M-N)$.
$\ket{\mathrm{D}_{j}^{b}}$ differs from the reference determinant by the single 
excitation $j \rightarrow b$. We decided to not sum over single excitations from 
the reference as in the case of self-consistent field reference determinants, 
Brillouin's theorem would mean that the weights would be (close to) zero.
Assuming the system is single referenced, we might 
assume that a doubly excited determinant might be second most important after 
the reference determinant. The sum is therefore over all possible double excited 
determinants trying to connect to a determinant slightly closer to the reference 
via a single excitation. For selecting $a$,
\begin{equation}
w_{a=a_{\sigma, \mathrm{sym.}},i,\mathrm{s}} = \frac{1}{n_{jb}} 
    \sum_{j=j_\mathrm{occ.ref.}, b=b_\mathrm{virt. ref.}} 
    \bra{\mathrm{D}_{j}^{b}} \hat{H} \ket{\mathrm{D}_{ij}^{ab}},
\end{equation}
is pre-computed where $i$ is now an occupied orbital in the current determinant which will have 
been selected before $w_{a=a_{\sigma, \mathrm{sym.}}i,\mathrm{s}}$ is needed. 
Given that the current determinant is not known at this stage, this is pre-computed 
for any orbital $i$. $a$ is then selected from the orbitals of allowed spin and 
symmetry for which $i \rightarrow a$ is valid. Alias tables are then pre-computed
for $w_{i',\mathrm{s}}$ and $w_{a=a_{\sigma, \mathrm{sym.}}i,\mathrm{s}}$.
\par When running the excitation generator, it is first decided whether a single or double excitation is attempted
with probability 
$p_\mathrm{single}$ or $1-p_\mathrm{single}$ respectively. If a single excitation was chosen, $i'$ is first selected in the 
reference frame from the occupied orbitals in the reference using the alias 
table constructed with weights $w_{i',\mathrm{s}}$. $i'$ is then mapped to the corresponding 
occupied orbital in the current determinant $i$ in the simulation frame.
Appendix \ref{app:mapping} explains the mapping between these two frames in detail.
\par Once $i$ is known, $a$ is selected using the pre-computed alias table 
with $w_{a=a_{\sigma, \mathrm{sym.}}i,\mathrm{s}}$. Of course, $a$ could be occupied. If 
that is the case, the excitation attempt was unsuccessful. Otherwise,
$i \rightarrow a$ is found and the generation probability is
\begin{equation}
\begin{split}
p_{\mathrm{gen, single}} = \\ p_\mathrm{single} \times 
    \frac{w_{i',\mathrm{s}}}{\sum_{i'=i'_\mathrm{occ.ref.}} w_{i',\mathrm{s}}} \times 
    \frac{w_{a=a_{\sigma, \mathrm{sym.}},i,\mathrm{s}}}{\sum_{a=a_{\sigma, \mathrm{sym.}}} 
    w_{a=a_{\sigma, \mathrm{sym.}},i,\mathrm{s}}}.
\end{split}
\end{equation}

\par For double excitations, four weight tables are pre-computed. For the 
selection of $i$ and $j$, heat bath weights are pre-computed, assuming the
reference determinant is fully occupied. Two orbitals $i'$ and $j'$ occupied
in the reference are found and then mapped to the actual determinant that is occupied.
For the virtual orbitals $a$ and $b$, alias tables based on Power--Pitzer weights
are pre-calculated for \textit{all} spinorbitals. Before selecting $a$, the actual $i$ is known
and can be substituted into pre-computed weights $\sqrt{|\braket{ia|ai}|}$ to find $a$.
 The memory cost is $\mathcal{O}(M^2)$.
No mapping is necessary for $a$ and $b$. Again, if $a$ or $b$ are occupied or $b$ is equal 
to $a$ or if there is not suitable orbital for $b$, the spawn attempt was 
unsuccessful. Double excitations with this excitation generator are explained in more detail
in appendix \ref{app:double}.
\par Overall, this is an excitation generator that is both weighted and scales as $\mathcal{O}(N_\mathrm{ex.})$
which does not scale with system size. The memory cost is also relatively small,
$\mathcal{O}(M^2)$.

\section{Results and Discussion}
\begin{figure}
\centering
        \begin{subfigure}[h]{8.5cm}
        \includegraphics[width=1.0\linewidth,keepaspectratio]{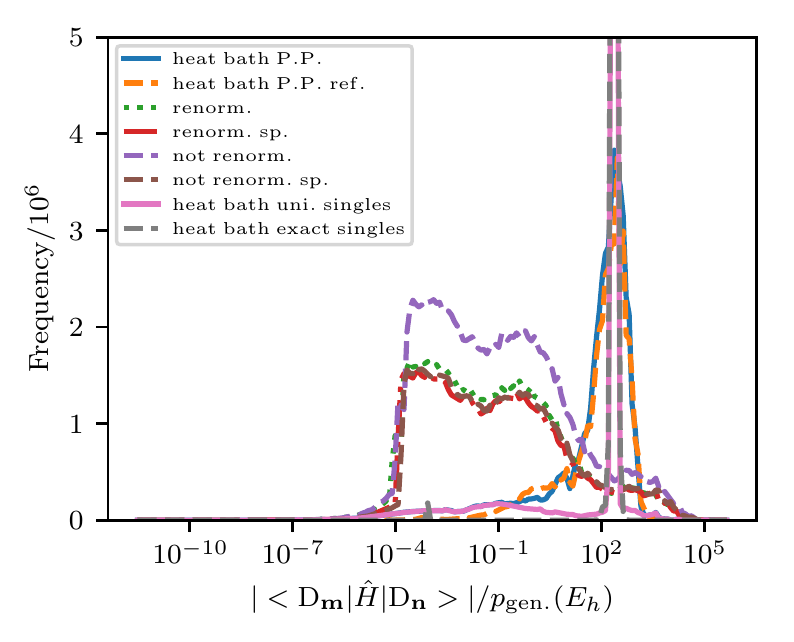}
        \end{subfigure}
        \newline
        \begin{subfigure}[h]{8.5cm}
        \includegraphics[width=1.0\linewidth,keepaspectratio]{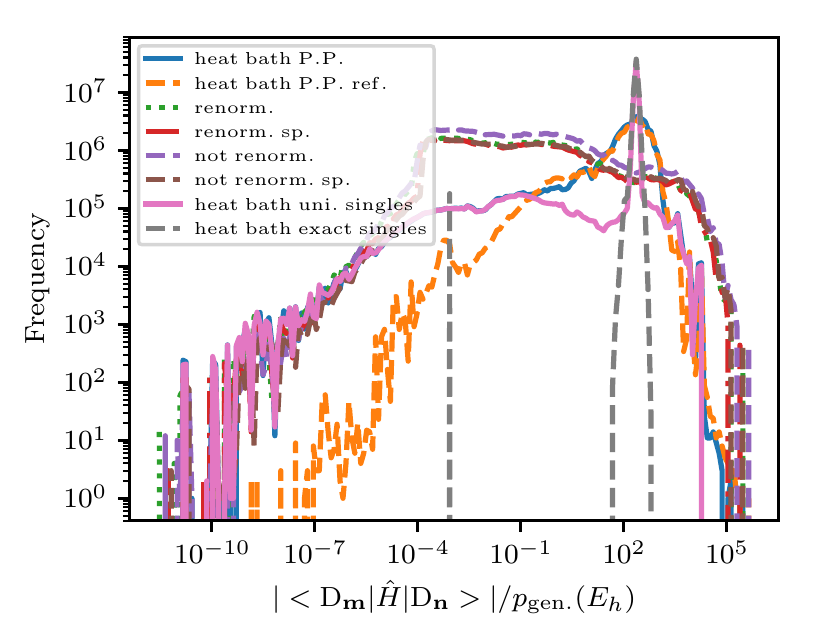}
        \end{subfigure}
        \newline
        \caption{Comparison of the histograms of $\frac{|\bra{\mathrm{D}_{\mathbf{n}}} \hat{H} \ket{\mathrm{D}_{\mathbf{m}}}|}{p_\mathrm{gen}}$ for various excitation generators.
        CCSD was performed on three water molecules in the cc-pVDZ basis using localized MOs.
        The values were logged for one Monte Carlo iteration. The size of the bins is logarithmic.
        \textit{Bottom} graph took the
        log of the frequency whereas the \textit{top} graph did not. They both show the same data.
        The frequency axis in the case is truncated in the \textit{top} graph. Most of them were restarted from the same calculation and then equilibrated before taking data. \textit{heat bath exact singles} was restarted from an
        equilibrated \textit{heat bath uniform singles} but not equilibrated since it is very slow.
        $\frac{|\bra{\mathrm{D}_{\mathbf{n}}} \hat{H} \ket{\mathrm{D}_{\mathbf{m}}}|}{p_\mathrm{gen}}=0$
        data is not shown which includes forbidden excitations.
        $p_\mathrm{single}$ was set to be the same when running which was corrected in post-processing to make
        the mean of finite $\frac{|\bra{\mathrm{D}_{\mathbf{n}}} \hat{H} \ket{\mathrm{D}_{\mathbf{m}}}|}{p_\mathrm{gen}}$
        for single and double excitations coincide which did not change $p_\mathrm{single}$ values by more than 30\%.}
        \label{fig:allhisto}
\end{figure}

To compare the effectiveness of the excitation generators discussed, water chains were then studied in a cc-pVDZ basis set \cite{Dunning1989} whose MOs have been localized.
Figure \ref{fig:allhisto} shows a histogram of 
$\frac{|\bra{\mathrm{D}_{\mathbf{m}}} \hat{H} 
\ket{\mathrm{D}_{\mathbf{n}}}|}{p_\mathrm{gen}}$ for three waters with the four uniform excitation
generators, the \textit{heat bath Power--Pitzer} excitation generator
(which had the sharpest peak out of the $\mathcal{O}(M)$/on-the-fly excitation generators),
the \textit{heat bath Power--Pitzer ref.} and the two \textit{heat bath} excitation generators that
do not suffer from bias. The ``original'' \textit{heat bath} excitation generator
was rejected by our bias test as it was not clear whether all allowed single excitations
can be created.
Considering a logarithmic scale in $\frac{|\bra{\mathrm{D}_{\mathbf{m}}} \hat{H} 
\ket{\mathrm{D}_{\mathbf{n}}}|}{p_\mathrm{gen}}$, the \textit{top} graph in figure
\ref{fig:allhisto} clearly shows that the uniform excitation generators produce a bigger
spread in $\frac{|\bra{\mathrm{D}_{\mathbf{m}}} \hat{H} 
\ket{\mathrm{D}_{\mathbf{n}}}|}{p_\mathrm{gen}}$ than weighted excitation generators
(\textit{Power--Pitzer} or \textit{heat bath}).

The \textit{heat bath} excitation generators
produce the sharpest peak. The \textit{heat bath uniform singles} excitation generator, that samples
single excitations uniformly, shares the main peak with the \textit{heat bath exact singles} excitation
generator, that samples single excitations exactly, but has a larger spread around that peak
caused by the uniform sampling of single excitations. The \textit{heat bath exact singles} excitation
generator produces two sharp peaks, both containing data from single excitations which were
treated exactly here. The reason why this is not one sharp peak is that in an ideal case
\begin{equation}
p_\mathrm{gen.} = \left|\frac{\bra{\mathrm{D}_{\mathbf{m}}} \hat{H} \ket{\mathrm{D}_{\mathbf{n}}}} {\sum_\mathbf{n} \bra{\mathrm{D}_{\mathbf{m}}} \hat{H} \ket{\mathrm{D}_{\mathbf{n}}}}\right|
\end{equation}
which means that
\begin{equation}
\frac{|\bra{\mathrm{D}_{\mathbf{m}}} \hat{H} \ket{\mathrm{D}_{\mathbf{n}}}|}{p_\mathrm{gen}} \approx \frac{1} {|\sum_\mathbf{n} \bra{\mathrm{D}_{\mathbf{m}}} \hat{H} \ket{\mathrm{D}_{\mathbf{n}}}|}
\end{equation}
in the case of an ideal excitation generator. This quantity depends on $\ket{\mathrm{D}_{\mathbf{n}}}$
and can therefore not be a constant in general
unless the selection step in the CCMC or FCIQMC algorithm is adapted as well.
Both \textit{heat bath} excitation generators here have a large memory scaling ($\mathcal{O}(M^4)$) and
\textit{heat bath exact singles} which produces the sharpest peak in the histogram has a computational scaling
of $\mathcal{O}(MN)$ which makes the \textit{heat bath exact singles} excitation generator not practical.

The main peak that the two \textit{Power--Pitzer} excitation generators produce is wider than
with the \textit{heat bath} excitation generators but it is significantly more compact that what
the uniform excitation generators give. The \textit{heat bath Power--Pitzer ref.} excitation generator
has a shorter tail on the low end but a slightly wider tail on the higher end. It has fewer than 250 events
in bins with bigger $\frac{|\bra{\mathrm{D}_{\mathbf{m}}} \hat{H} 
\ket{\mathrm{D}_{\mathbf{n}}}|}{p_\mathrm{gen}}$ than the highest bin that has an event with the
\textit{heat bath uniform singles} excitation generator. The \textit{heat bath Power--Pitzer} excitation generator
has fewer than 90 events above the bin with highest $\frac{|\bra{\mathrm{D}_{\mathbf{m}}} \hat{H} 
\ket{\mathrm{D}_{\mathbf{n}}}|}{p_\mathrm{gen}}$ in the \textit{heat bath uniform singles} case.
\begin{table}
	\caption{Fraction of non-zero allowed spawn events, both with respect to total number
	of spawn events. The latter represents the spawn events depicted in figure \ref{fig:allhisto}.
	The \textit{renorm.} and \textit{renorm. spin} have been combined to \textit{renorm.} and similarly
	for  \textit{not. renorm.}. \textit{P.--P.} means \textit{Power--Pitzer} and \textit{heat b.} is \textit{heat bath}.}

		\begin{tabular}{|l|l|}
		\hline
			& $\frac{\mathrm{\#allowed}\ \mathrm{non-zero}}{\mathrm{\#total}}$ events \\
			\hline
			\textit{heat b. P.--P. ref.} &  0.66--0.67 \\
			\textit{heat b. P.--P.} &  0.68--0.69 \\
			\textit{heat b. uniform singles} &  0.72 \\
			\textit{heat b. exact singles} &  0.72 \\
			\textit{renorm.} &  0.68--0.72 \\
			\textit{not. renorm.} &  0.54--0.57 \\\hline
		\end{tabular}

	\label{tab:zerodisallowedtot}
\end{table} 
The number of finite $\frac{|\bra{\mathrm{D}_{\mathbf{m}}} \hat{H} \ket{\mathrm{D}_{\mathbf{n}}}|}{p_\mathrm{gen}}$,
allowed events are shown in table \ref{tab:zerodisallowedtot}.
The weighted excitation generators have similar fractions of allowed non-zero events and
the \textit{heat bath Power--Pitzer ref.} excitation generator has the lower computational scaling
compared to \textit{heat bath Power--Pitzer} and %--- in principle ---
 the \textit{heat bath uniform singles} excitation generator. It also does not have the prohibitively large memory scaling of
the \textit{heat bath uniform singles} excitation generator.

Next, we move away from abstract performance considerations and consider how the different excitation generators
affect the
the efficiency (as described by Holmes et al. \cite{Holmes2016a}), inefficiency\cite{Vigor2016},
and the position of the shoulder\cite{Spencer2016}
which are all measures of the difficulty of the calculation.  The efficiency $\eta$ is defined as $\eta=1/(\sigma_E^2 T)$,
where $\sigma_E$ is the statistical error in the energy (here projected energy) and $T$ is
the computational time taken to achieve error bar $\sigma_E$. Note that this does not include convergence wall-time.
We have found $T$ to be highly dependent
on implementation so $\eta$ must be considered carefully. We also consider the (theoretical) algorithmic computational scaling in mind and
the inefficiency $a$ as defined by Vigor et al. \cite{Vigor2016},
$a = \sigma_E \sqrt{\delta \tau N_\mathrm{it.} \langle N_\mathrm{p}\rangle}$ where 
$N_\mathrm{it.}$ is the number of iterations considered in the blocking analysis 
and $\langle N_\mathrm{p}\rangle$ is the mean number of Monte Carlo particles. When 
estimating the error in the efficiency and inefficiency, we ignore the correlation in the 
numerator and denominator of the $E_\mathrm{proj.}$, so giving an upper bound on the error.
In a non-initiator calculation, the shoulder is a feature in a graph of
total excip population against ratio of total population to population on the
reference at the point when enough excips are in the calculation to converge to the
correct wavefunction. After that point, the population controlling shift can be varied and
data can be taken. It is therefore a measure of how many excips have to be in the calculation.

We have varied the 
shift damping automatically to reduce the variance of the projected energy.
\footnote{Feature implemented by Charles Scott.}

\subsection{Coupled Cluster Monte Carlo}
\begin{figure}
\centering
	\begin{subfigure}[h]{\linewidth}
        \includegraphics[width=1.0\linewidth,keepaspectratio]{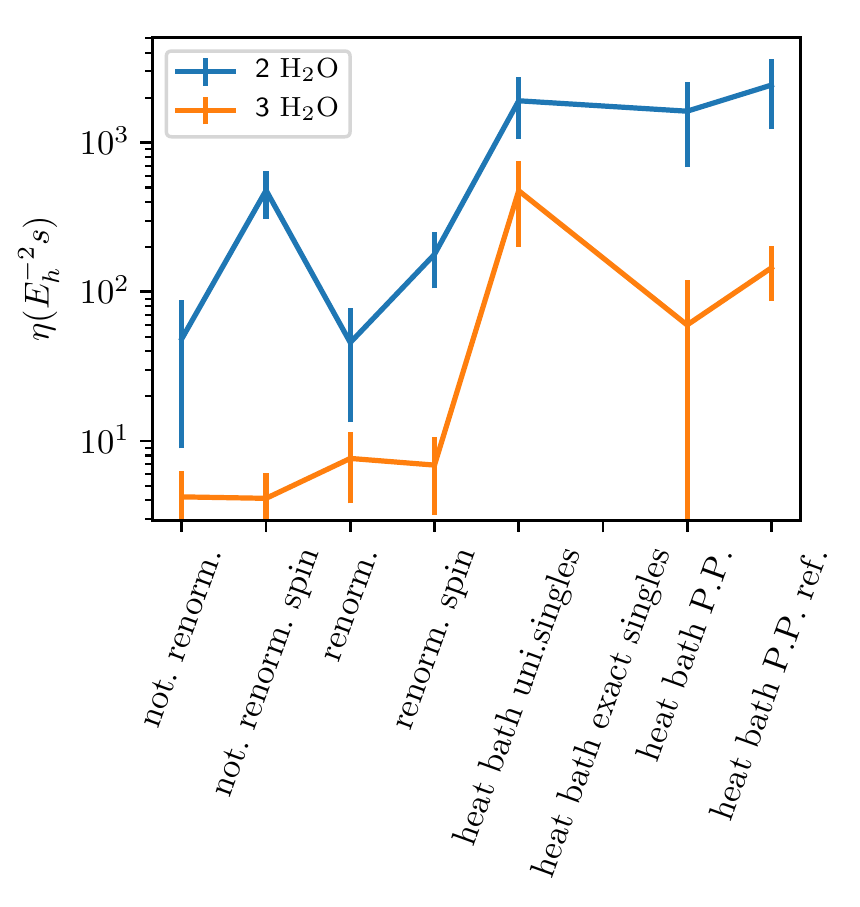}%
	\end{subfigure}
		\begin{subfigure}[h]{\linewidth}
			\includegraphics[width=1.0\linewidth,keepaspectratio]{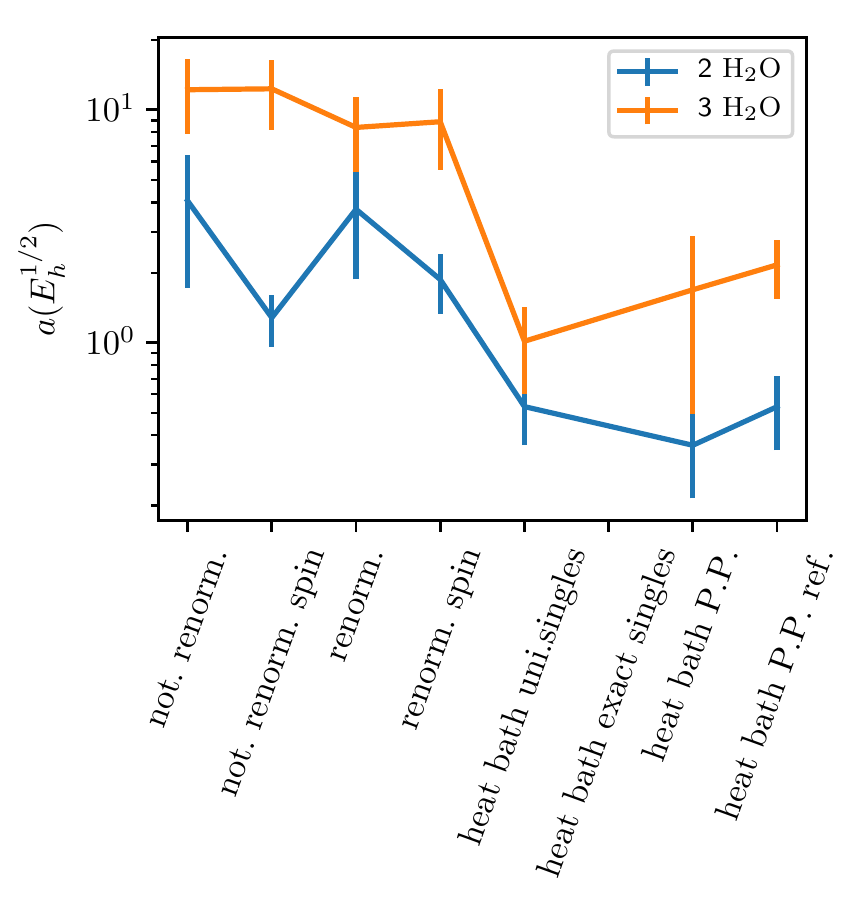}%
		\end{subfigure}%
	\caption{Efficiency $\eta$ (\textit{top}) and inefficiency $a$ (\textit{bottom}) for
	chains of two or three water molecules in a cc-pVDZ basis run with CCSD using localised MOs. Error bars neglect
	the covariance between numerator and denominator errors in the projected energy and are over-estimates.
        P.P. stands for \textit{Power--Pitzer}. The \textit{heat bath exact singles} excitation generator was too slow
	for data to be taken. The different excitation generators were run under the same 
	conditions with the same time step etc. Only the target population varies between the water dimer and trimer calculation.
	The starting iteration for heat bath P.P. was found such that 3 reblocks could be used.}
			\label{fig:effccsdccpV}
\end{figure}

All coupled cluster calculations are non-initiator \cite{Cleland2010,Spencer2016}.
Figure \ref{fig:effccsdccpV} shows the efficiency and inefficiency for chains of two and three waters in the cc-pVDZ basis performing CCSD with localised molecular orbitals. To get small enough error bars on efficiency and inefficiency, the systems to study cannot be too large. The \textit{heat bath uniform singles} and the \textit{heat bath Power--Pitzer ref.} excitation generators assume that the number of occupied orbitals is small relative to the number of total orbitals, which reflects a realistic calculation, so our basis set cannot be too small.
The error bars efficiency and inefficiency have been estimated by neglecting the covariance between numerator and the denominator errors in the projected energy to give an upper bound.
The \textit{heat bath exact singles} excitation generator is so slow that
it was not possible to take sufficient data with it to produce results. The trend is that the weighted excitation generators are more efficient and
less inefficient than the uniform ones. This becomes more noticeable in the larger system. As expected, modelling three waters is less efficient and more inefficient than two, the difference being more distinct with the uniform excitation generators. The \textit{heat bath uniform singles} excitation performs best out of the weighted ones which is expected due to the same computational scaling as \textit{heat bath Power--Pitzer ref.} but a more favourable scaling than \textit{heat bath Power--Pitzer} while using well approximated weights
for double excitations. For the trimer calculation, it was difficult to block the data of \textit{heat bath Power--Pitzer} and to achieve convergence due to its scaling with the number of spinorbitals.
\par
\begin{figure}
	\centering
	\begin{subfigure}[h]{\linewidth}
		\includegraphics[width=1.0\linewidth,keepaspectratio]{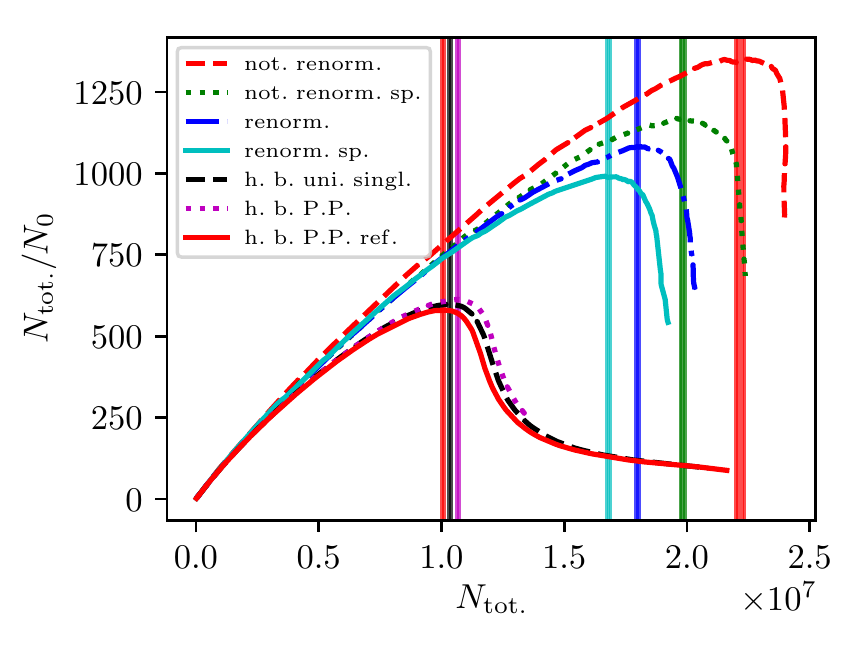}
	\end{subfigure}
	\caption{Shoulder plots for two localised waters in a cc-pVDZ basis with CCSDT with various excitation generators. P.P. stands for \textit{Power--Pitzer}, h.b. for \textit{heat bath} and sp. for \textit{spin}. The different excitation generators were run under similar 
		conditions with the same time step etc. The weighted excitations generators started varying the shift after
		a total population of 20 million whereas the uniform ones did not vary the shift.
		The vertical lines 
		represent the ``shoulder height'', the position of the maximum 
		plus/minus of a standard deviation. To determine the shoulder 
		position, the mean and standard error of the mean of the 10 highest 
		data points were taken.}
			\label{figshoulder}
\end{figure}

 Next, we consider shoulder heights. Figure \ref{figshoulder} shows shoulder plots were the difference in shoulder positions between the
excitation generators is very clear.
The weighted excitation generators again perform best. Their shoulders are significantly lower than those of uniform excitation generators, by
a factor of just under 2. Of those studied, the \textit{heat bath Power--Pitzer ref.} excitation generator has the lowest shoulder.
\par These results show that the weighted excitation generators perform better than the uniform ones.
The \textit{heat bath Power--Pitzer ref.} excitation generator can scale independently of system
size computationally which puts it at a clear advantage over the \textit{heat bath Power--Pitzer} excitation generator.
It also has a reduced memory scaling when comparing it to the \textit{heat bath} excitation generators which is significant at bigger systems.

\subsection{Full Configuration Interaction Quantum Monte Carlo}
Next, we turn to FCIQMC. The water chain with two waters in cc-pVDZ basis with localized MOs was considered with initiator FCIQMC.
The (in--)efficiencies are determined at one point in the initiator curve (total population against energy).
All calculations were started with the same parameters, which included the population at which the shift started varying,
and so the eventual equilibrated population of the system then equilibrated is dependent upon the excitation generator.
Blooms did happen.
For uniform excitation generators it was over $10^7$ particles, for the
weighted ones $5.6\times10^6$.  Use of a larger population may lead to a decrease in measured inefficiency\cite{Vigor2016}, so the results from the uniform excitation generators should be regarded as lower bounds for inefficiency.

\begin{figure}
	\centering
	\begin{subfigure}[h]{\linewidth}
		\includegraphics[width=1.0\linewidth,keepaspectratio]{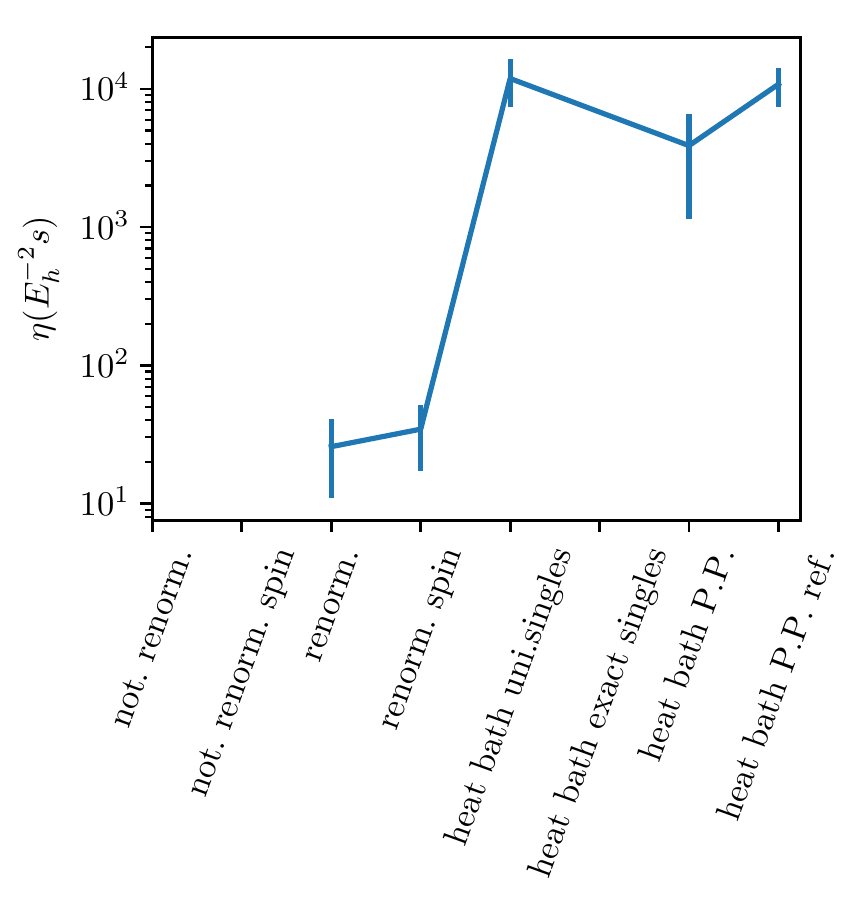}
	\end{subfigure}
	\begin{subfigure}[h]{\linewidth}
		\includegraphics[width=1.0\linewidth,keepaspectratio]{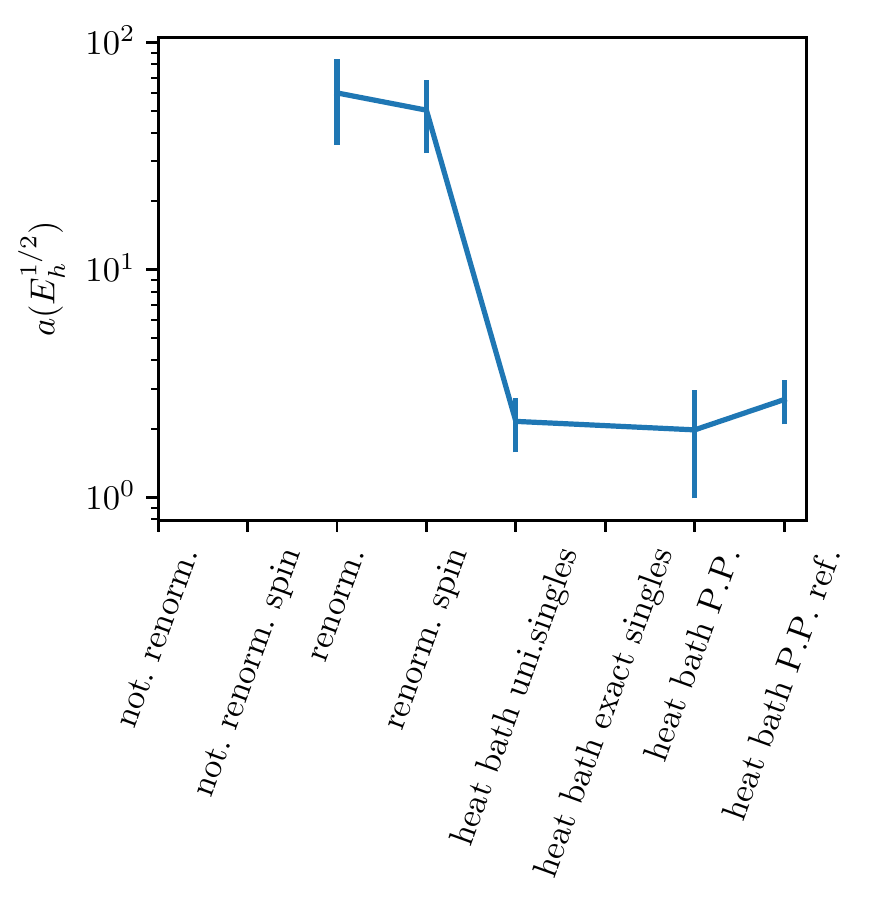}
	\end{subfigure}
	\caption{Efficiency $\eta$ (\textit{top}) and inefficiency $a$ (\textit{bottom}) for a
		chain of two water molecules in cc-pVDZ basis using localized MOs run with initiator FCIQMC. Error bars neglect
	the covariance between numerator and denominator errors in the projected energy and are over-estimates.
		P.P. stands for \textit{Power--Pitzer}. The \textit{heat bath exact singles} excitation generator was too slow
		for data to be taken. The different excitation generators were run under the same 
		conditions with the same time step etc. The spawning arrays of the \textit{not. renorm.} excitation generators
		ran out of memory so the space to store the spawned walkers would need to be increased for those results.}
			\label{fig:effifciqmcccpVDZ2}
\end{figure}
Figure \ref{fig:effifciqmcccpVDZ2} shows the efficiency and inefficiency for that system. The weighted excitation generators perform comparably among themselves and all outperform the uniform ones.
\textit{heat bath Power--Pitzer ref.} and \textit{heat bath uniform singles} both scale linearly in the number of electrons when using FCIQMC.
Holmes et al. \cite{Holmes2016a} describe ways to reduce the memory cost by 
considering spins (we just store zeroes instead of considering the spin when 
selecting) or by not storing all the weight to select $b$ for example.
We have used double precision for the weights. However, even if our implementation is not optimal, 
it is clear that the \textit{heat bath} excitation generators hit a memory 
ceiling with big systems significantly earlier than the \textit{heat bath Power--Pitzer ref.} excitation 
generators.
\par This shows that \textit{heat bath Power--Pitzer ref.} is an efficient excitation generator with a low shoulder
that can be used in CCMC and FCIQMC as a weighted excitation generator with low computational and memory cost.

\section{Conclusion}
We have shown that the \textit{heat bath Power--Pitzer ref.} excitation generator
combines the advantages of \textit{heat bath} excitation generators, which are relatively fast and
use good weights but struggle with a significant memory cost and a possible bias,
and the excitation generators that approximate heat bath weights by inequalities which are
calculated on-the-fly reducing the memory scaling but scaling prohibitively computationally in big systems.
The  \textit{heat bath Power--Pitzer ref.} excitation generator has at worst a low computational order and can be implemented with computational cost independent of system size in coupled cluster with a low memory cost.

\begin{acknowledgements}
We thank Prof. Ali Alavi and Dr. Pablo L\'opez R\'ios for helpful discussions. 
    Supporting research data and further information can be found at 
    \url{doi.org/XXXX}. V.A.N. would like to acknowledge the EPSRC Centre for 
    Doctoral Training in Computational Methods for Materials Science for funding 
    under grant number EP/L015552/1 and A.J.W.T. thanks the Royal Society for a 
    University Research Fellowship under grants UF110161 and UF160398. This work 
    used the ARCHER UK National Supercomputing Service (http://www.archer.ac.uk) 
    and the UK Research Data Facility \\
    (http://www.archer.ac.uk/documentation/rdf-guide) under ARCHER Leadership 
    project with grant number e507.
\end{acknowledgements}

\appendix
\section{\label{app:furtheruni} Further Uniform Excitation Generators}
\par In the case of a double excitation, Hamiltonian matrix elements tend to be 
bigger if $i$ and $j$ do not have parallel spins. This is because following 
Slater-Condon rules \cite{Slater1929,Condon1930}, the Hamiltonian 
matrix element is reduced to a sum of two terms of opposite sign in the case of 
parallel spins ($\braket{ij|ab} - \braket{ij|ba}$, see later section for definition) and one term if the spins are 
not parallel ($\braket{ij|ab}$). It might therefore be advisable to 
select anti-parallel spin electrons with a greater probability than parallel 
$ij$. Alavi, Booth and others\cite{Booth2014}\footnote{Personal Communication with Ali Alavi and Pablo L\'opez R\'ios. This is also
implemented in NECI \url{https://github.com/ghb24/NECI_STABLE}.} had the idea of determining whether spins are 
antiparallel or parallel first when selecting $i$ and $j$. The \textit{no. 
renorm. spin} and \textit{renorm. spin} excitation generators are modifications 
of \textit{no. renorm.} and \textit{renorm.} excitation generators, where 
instead of finding $i$ and $j$ as a pair from the set of occupied orbitals, it is first decided
whether they should have parallel spins or not. With probability 
$p_\mathrm{parallel}$, $ij$ are either selected as a pair from the set of occupied 
$\alpha$ (probability $\frac{N_\alpha}{N}$) or from the set of occupied $\beta$ 
orbitals (probability $1-\frac{N_\alpha}{N}=\frac{N_\beta}{N}$) where $N_\alpha$ 
and $N_\beta$ are the number of $\alpha$ and $\beta$ electrons respectively.
 This can lead to forbidden 
excitations followed by failed spawning attempts if there is only one electron 
of one type of spin. Here, $p_\mathrm{parallel}$ is set as the fraction of $H_{ijab}$ where $i$ 
and $j$ have parallel spins.

\section{\label{app:mapping} Mapping spinorbitals in \textit{heat bath Power--Pitzer ref.} excitation generator}
\begin{figure}
        \centering
        \includegraphics[width=8cm,keepaspectratio,trim={0 15cm 3cm 
        0}]{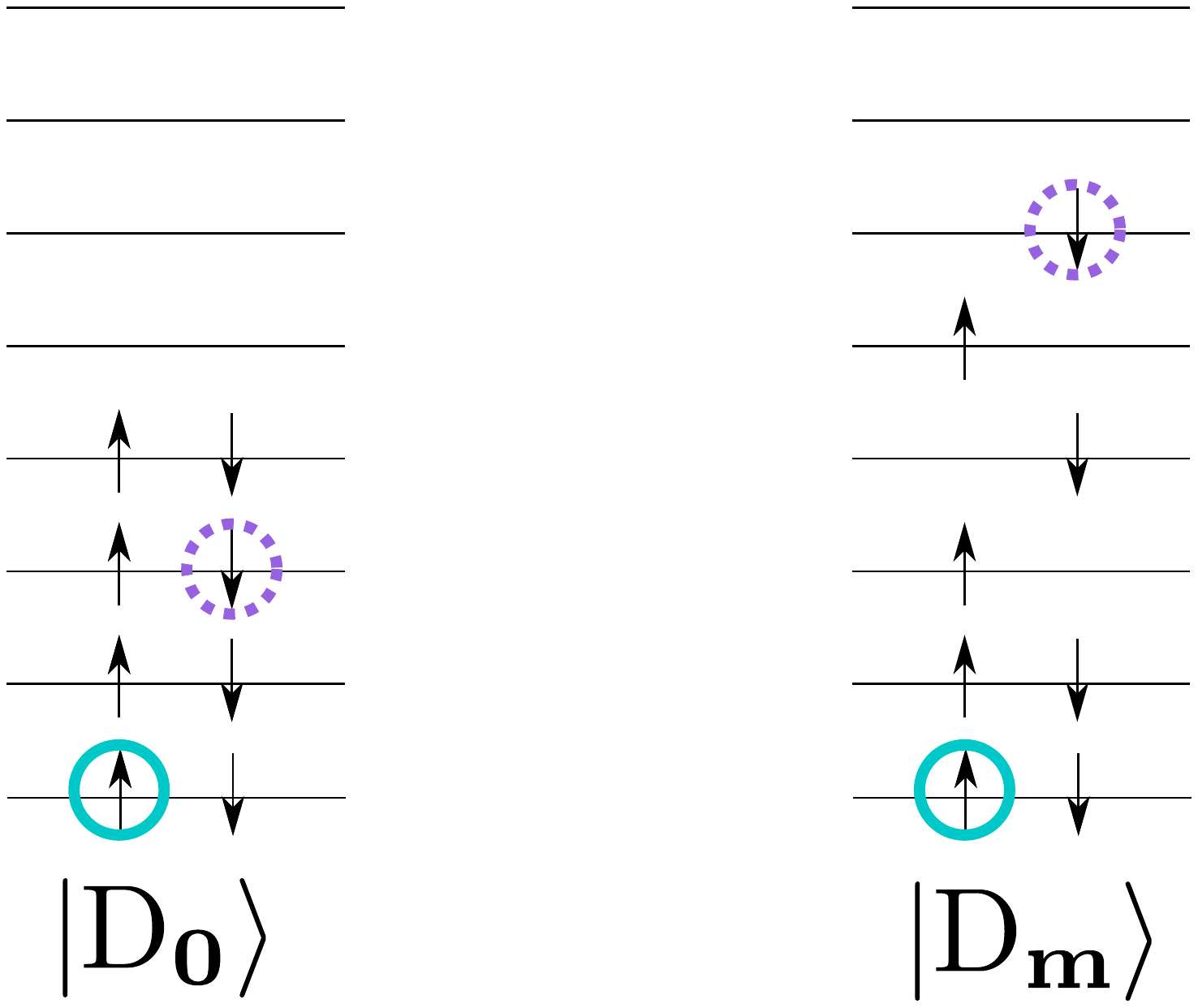}
        \caption{Selecting $i$ and $j$ with \textit{heat bath Power--Pitzer ref.} 
         excitation generator for a double excitation. First 
        $i'$ is selected, occupied in the reference determinant 
        $\ket{\mathrm{D}_\mathbf{0}}$ and translated to $i$, occupied in the 
        current determinant, $\ket{\mathrm{D}_\mathbf{m}}$. $i'$ and $i$ are 
        shown with light blue solid circles. In this case, $i' = i$. Then $j'$ is found and translated to $j$.
        As $j'$ is not occupied in 
        $\ket{\mathrm{D}_\mathbf{m}}$, it is mapped to the next orbital of the same 
        spin occupied in $\ket{\mathrm{D}_\mathbf{m}}$ but not in 
        $\ket{\mathrm{D}_\mathbf{0}}$. $j'$ and $j$ are shown with dashed purple 
        circles. Here $j' \neq j$.}
        \label{fig:mapping}
\end{figure}

In HANDE, there is a list of orbitals that are occupied in the reference, usually approximately ordered by 
one electron energies, and there is an equivalent ordered list 
with orbitals occupied in the current determinant $\ket{\mathrm{D}_\mathbf{m}}$. 
Every time $\ket{\mathrm{D}_\mathbf{m}}$ is changed, two new (energy ordered) lists $RD$ and $CD$
are created, one ($RD$) containing 
all orbitals that are occupied in the reference but not in 
$\ket{\mathrm{D}_\mathbf{m}}$ and another list ($CD$) of the same size with all 
orbitals occupied in $\ket{\mathrm{D}_\mathbf{m}}$ but not the reference 
determinant. Orbitals with the same positions in these two 
lists are made to have the same spin by swapping orbitals in the list $CD$ if necessary. 
If necessary, orbitals are translated by a one-to-one mapping between these two lists.
If $i'$ is not only occupied in the reference but in 
$\ket{\mathrm{D}_\mathbf{m}}$ as well, $i' = i$. If not, the position 
$i'$ has in list $RD$ is translated to the orbital with the same position in 
list $CD$. 
Figure \ref{fig:mapping} shows the translation of 
$i$ and $j$ in a double excitation in the two frames of reference pictorially.
Note that this is the only part of the excitation
generator that is not $\mathcal{O}(1)$ but $\mathcal{O}(N)$, mainly arising due to the creation of the
two lists. The computational cost
is reduced to $\mathcal{O}(1)$ if a determinant is reused. Alternatively, if, as mentioned
previously, each excitor is not represented by a determinant but rather the lists
$RD$ or $CD$ from the beginning the scaling is reduced to $\mathcal{O}(N_\mathrm{ex.})$
which is the cost of finding the correct mapping from one list to the other.

\section{\label{app:double} Details of double excitations in the \textit{heat bath Power--Pitzer ref.} excitation generator}
Again, orbitals $i'j'$ are part of the reference frame, where the reference determinant is occupied, and
$ij$ are the equivalent spinorbitals in the actual frame, where the actual determinant we are exciting from is occupied.
$i'j'$ are first found in the reference frame using heat bath weights and then they are mapped to the actual frame
as described in appendix \ref{app:mapping}. $ab$ are found with Power--Pitzer weights in the actual frame. All weights
are pre-computed. This appendix describes the details of generating the double excitation.
For $i'$, the pre-computed weights are
\begin{equation}
w_{i',\mathrm{d}} = \sum_{j'=j_{\mathrm{occ.ref.},\neq i'},a_{\neq \{i',j'\}},b_{\neq 
    \{i',j',a\}}}H_{i'j'ab}
\end{equation}
$i'$ is selected from the set of occupied orbitals in the 
reference with a sum over $j'$, the set of occupied orbitals in the reference 
other than $i'$. $a$ and $b$ out of the set of all orbitals (not 
just virtual) are summed over, provided they don't equal $i'$, $j'$ or each other. For $j'$,
\begin{equation}
w_{j'i,\mathrm{d}} = \sum_{a_{\neq \{i,j'\}},b_{\neq \{i,j',a\}}}H_{ij'ab}
\end{equation}
is pre-calculated which is of order $\mathcal{O}(NM)$. For both $w_{i',\mathrm{d}}$ and $w_{j'i,\mathrm{d}}$, 
a minimum weight is set in case the total weight for selecting $i'$ 
or $j'$ respectively in the reference frame is zero but selecting the equivalent 
$i$ and $j$ in the simulation frame would be allowed.
\par To select $a$ and $b$, Power--Pitzer weights are pre-calculated. For $a$,
\begin{equation}
w_{a,i,\mathrm{d}} = \sqrt{|\braket{ia|ai}|}
\end{equation}
where $w_{a,i,\mathrm{d}}$ is zero if $i = a$. $ia$ are from the set of all spinorbitals
and $a$ is restricted to the set of
the same spin as $i$. The memory cost is simply $\mathcal{O}(M^2)$. Similarly, 
for $b$
\begin{equation}
w_{b,j,\mathrm{sym.},\mathrm{d}} = \sqrt{|\braket{jb|bj}|}
\end{equation}
where $w_{b,j,\mathrm{d}} = 0$ if $b = j$ and $b$ is from the set of all spinorbitals 
with the same spin as $j$. $w_{b,j,\mathrm{d}}$ are arranged in such a way that  
$b$'s of the required symmetry later can readily be looked up. Alias 
tables for all these weights for single and double excitations are pre-computed.
\par In the case of a double excitation, first $i'$, an occupied orbital in 
the reference frame, is selected using $w_{i',\mathrm{d}}$. $i' \rightarrow i$ is mapped to an 
occupied orbital $i$ in the simulation frame if required. Then, $j'$ 
is found using the pre-computed alias table for $w_{j'i,\mathrm{d}}$ and map $j' \rightarrow j$ if 
needed. $i$ and $j$ are ordered so that $j$ has a higher or equal index in the determinant list
as $i$. 
Using $i$ and $w_{a,i,\mathrm{d}}$, $a$ is found using pre-computed alias tables out of 
all spinorbitals with the same spin as $i$. If $a$ is occupied, the spawn 
attempt was unsuccessful. The symmetry that $b$ should have is then determined and 
using the pre-calculated alias tables for $w_{b,j,\mathrm{sym.},\mathrm{d}}$ which give 
us a $b$ of the correct symmetry (and spin), $b$ is found from the set of all 
spinorbitals with required spin and symmetry. Again, if $b$ is occupied or equal 
to $a$ or if there is not suitable orbital for $b$, the spawn attempt was 
unsuccessful.

% Create the reference section using BibTeX:
%\bibliography{library}

\begin{thebibliography}{73}%
\makeatletter
\providecommand \@ifxundefined [1]{%
 \@ifx{#1\undefined}
}%
\providecommand \@ifnum [1]{%
 \ifnum #1\expandafter \@firstoftwo
 \else \expandafter \@secondoftwo
 \fi
}%
\providecommand \@ifx [1]{%
 \ifx #1\expandafter \@firstoftwo
 \else \expandafter \@secondoftwo
 \fi
}%
\providecommand \natexlab [1]{#1}%
\providecommand \enquote  [1]{``#1''}%
\providecommand \bibnamefont  [1]{#1}%
\providecommand \bibfnamefont [1]{#1}%
\providecommand \citenamefont [1]{#1}%
\providecommand \href@noop [0]{\@secondoftwo}%
\providecommand \href [0]{\begingroup \@sanitize@url \@href}%
\providecommand \@href[1]{\@@startlink{#1}\@@href}%
\providecommand \@@href[1]{\endgroup#1\@@endlink}%
\providecommand \@sanitize@url [0]{\catcode `\\12\catcode `\$12\catcode
  `\&12\catcode `\#12\catcode `\^12\catcode `\_12\catcode `\%12\relax}%
\providecommand \@@startlink[1]{}%
\providecommand \@@endlink[0]{}%
\providecommand \url  [0]{\begingroup\@sanitize@url \@url }%
\providecommand \@url [1]{\endgroup\@href {#1}{\urlprefix }}%
\providecommand \urlprefix  [0]{URL }%
\providecommand \Eprint [0]{\href }%
\providecommand \doibase [0]{http://dx.doi.org/}%
\providecommand \selectlanguage [0]{\@gobble}%
\providecommand \bibinfo  [0]{\@secondoftwo}%
\providecommand \bibfield  [0]{\@secondoftwo}%
\providecommand \translation [1]{[#1]}%
\providecommand \BibitemOpen [0]{}%
\providecommand \bibitemStop [0]{}%
\providecommand \bibitemNoStop [0]{.\EOS\space}%
\providecommand \EOS [0]{\spacefactor3000\relax}%
\providecommand \BibitemShut  [1]{\csname bibitem#1\endcsname}%
\let\auto@bib@innerbib\@empty
%</preamble>
\bibitem [{\citenamefont {Coester}\ and\ \citenamefont
  {K{\"{u}}mmel}(1960)}]{Coester1960}%
  \BibitemOpen
  \bibfield  {author} {\bibinfo {author} {\bibfnamefont {F.}~\bibnamefont
  {Coester}}\ and\ \bibinfo {author} {\bibfnamefont {H.}~\bibnamefont
  {K{\"{u}}mmel}},\ }\href {\doibase 10.1016/0029-5582(60)90140-1} {\bibfield
  {journal} {\bibinfo  {journal} {Nucl. Phys.}\ }\textbf {\bibinfo {volume}
  {17}},\ \bibinfo {pages} {477} (\bibinfo {year} {1960})}\BibitemShut
  {NoStop}%
\bibitem [{\citenamefont {{\v{C}}i{\v{z}}ek}(1966)}]{Cizek1966}%
  \BibitemOpen
  \bibfield  {author} {\bibinfo {author} {\bibfnamefont {J.}~\bibnamefont
  {{\v{C}}i{\v{z}}ek}},\ }\href {\doibase 10.1063/1.1727484} {\bibfield
  {journal} {\bibinfo  {journal} {J. Chem. Phys.}\ }\textbf {\bibinfo {volume}
  {45}},\ \bibinfo {pages} {4256} (\bibinfo {year} {1966})}\BibitemShut
  {NoStop}%
\bibitem [{\citenamefont {{\v{C}}i{\v{z}}ek}\ and\ \citenamefont
  {Paldus}(1971)}]{Cizek1971}%
  \BibitemOpen
  \bibfield  {author} {\bibinfo {author} {\bibfnamefont {J.}~\bibnamefont
  {{\v{C}}i{\v{z}}ek}}\ and\ \bibinfo {author} {\bibfnamefont {J.}~\bibnamefont
  {Paldus}},\ }\href {\doibase 10.1002/qua.560050402} {\bibfield  {journal}
  {\bibinfo  {journal} {Int. J. Quantum Chem.}\ }\textbf {\bibinfo {volume}
  {5}},\ \bibinfo {pages} {359} (\bibinfo {year} {1971})}\BibitemShut {NoStop}%
\bibitem [{\citenamefont {Bartlett}\ and\ \citenamefont
  {Musia{\l}}(2007)}]{Bartlett2007}%
  \BibitemOpen
  \bibfield  {author} {\bibinfo {author} {\bibfnamefont {R.~J.}\ \bibnamefont
  {Bartlett}}\ and\ \bibinfo {author} {\bibfnamefont {M.}~\bibnamefont
  {Musia{\l}}},\ }\href {\doibase 10.1103/RevModPhys.79.291} {\bibfield
  {journal} {\bibinfo  {journal} {Rev. Mod. Phys.}\ }\textbf {\bibinfo {volume}
  {79}},\ \bibinfo {pages} {291} (\bibinfo {year} {2007})}\BibitemShut
  {NoStop}%
\bibitem [{\citenamefont {Lee}\ and\ \citenamefont {Scuseria}(1995)}]{Lee1995}%
  \BibitemOpen
  \bibfield  {author} {\bibinfo {author} {\bibfnamefont {T.~J.}\ \bibnamefont
  {Lee}}\ and\ \bibinfo {author} {\bibfnamefont {G.~E.}\ \bibnamefont
  {Scuseria}},\ }in\ \href {\doibase 10.1007/978-94-011-0193-6_2} {\emph
  {\bibinfo {booktitle} {Quantum Mech. Electron. Struct. Calc. with Chem.
  Accuracy}}}\ (\bibinfo  {publisher} {Springer Netherlands},\ \bibinfo
  {address} {Dordrecht},\ \bibinfo {year} {1995})\ pp.\ \bibinfo {pages}
  {47--108}\BibitemShut {NoStop}%
\bibitem [{\citenamefont {Thom}(2010)}]{Thom2010}%
  \BibitemOpen
  \bibfield  {author} {\bibinfo {author} {\bibfnamefont {A.~J.~W.}\
  \bibnamefont {Thom}},\ }\href {\doibase 10.1103/PhysRevLett.105.263004}
  {\bibfield  {journal} {\bibinfo  {journal} {Phys. Rev. Lett.}\ }\textbf
  {\bibinfo {volume} {105}},\ \bibinfo {pages} {263004} (\bibinfo {year}
  {2010})}\BibitemShut {NoStop}%
\bibitem [{\citenamefont {Spencer}\ and\ \citenamefont
  {Thom}(2016)}]{Spencer2016}%
  \BibitemOpen
  \bibfield  {author} {\bibinfo {author} {\bibfnamefont {J.~S.}\ \bibnamefont
  {Spencer}}\ and\ \bibinfo {author} {\bibfnamefont {A.~J.~W.}\ \bibnamefont
  {Thom}},\ }\href {\doibase http://dx.doi.org/10.1063/1.4942173} {\bibfield
  {journal} {\bibinfo  {journal} {J. Chem. Phys.}\ }\textbf {\bibinfo {volume}
  {144}},\ \bibinfo {pages} {084108} (\bibinfo {year} {2016})},\ \Eprint
  {http://arxiv.org/abs/1511.05752} {arXiv:1511.05752} \BibitemShut {NoStop}%
\bibitem [{\citenamefont {Franklin}\ \emph {et~al.}(2016)\citenamefont
  {Franklin}, \citenamefont {Spencer}, \citenamefont {Zoccante},\ and\
  \citenamefont {Thom}}]{Franklin2016}%
  \BibitemOpen
  \bibfield  {author} {\bibinfo {author} {\bibfnamefont {R.~S.~T.}\
  \bibnamefont {Franklin}}, \bibinfo {author} {\bibfnamefont {J.~S.}\
  \bibnamefont {Spencer}}, \bibinfo {author} {\bibfnamefont {A.}~\bibnamefont
  {Zoccante}}, \ and\ \bibinfo {author} {\bibfnamefont {A.~J.~W.}\ \bibnamefont
  {Thom}},\ }\href {\doibase 10.1063/1.4940317} {\bibfield  {journal} {\bibinfo
   {journal} {J. Chem. Phys.}\ }\textbf {\bibinfo {volume} {144}},\ \bibinfo
  {pages} {044111} (\bibinfo {year} {2016})},\ \Eprint
  {http://arxiv.org/abs/1511.08129} {arXiv:1511.08129} \BibitemShut {NoStop}%
\bibitem [{\citenamefont {Scott}\ and\ \citenamefont
  {Thom}(2017)}]{Scott2017b}%
  \BibitemOpen
  \bibfield  {author} {\bibinfo {author} {\bibfnamefont {C.~J.~C.}\
  \bibnamefont {Scott}}\ and\ \bibinfo {author} {\bibfnamefont {A.~J.~W.}\
  \bibnamefont {Thom}},\ }\href {\doibase 10.1063/1.4991795} {\bibfield
  {journal} {\bibinfo  {journal} {J. Chem. Phys.}\ }\textbf {\bibinfo {volume}
  {147}},\ \bibinfo {pages} {124105} (\bibinfo {year} {2017})},\ \Eprint
  {http://arxiv.org/abs/1706.07017} {arXiv:1706.07017} \BibitemShut {NoStop}%
\bibitem [{\citenamefont {Spencer}\ \emph {et~al.}(2018)\citenamefont
  {Spencer}, \citenamefont {Neufeld}, \citenamefont {Vigor}, \citenamefont
  {Franklin},\ and\ \citenamefont {Thom}}]{Spencer2018}%
  \BibitemOpen
  \bibfield  {author} {\bibinfo {author} {\bibfnamefont {J.~S.}\ \bibnamefont
  {Spencer}}, \bibinfo {author} {\bibfnamefont {V.~A.}\ \bibnamefont
  {Neufeld}}, \bibinfo {author} {\bibfnamefont {W.~A.}\ \bibnamefont {Vigor}},
  \bibinfo {author} {\bibfnamefont {R.~S.~T.}\ \bibnamefont {Franklin}}, \ and\
  \bibinfo {author} {\bibfnamefont {A.~J.~W.}\ \bibnamefont {Thom}},\ }\href
  {https://arxiv.org/pdf/1807.03749.pdf http://arxiv.org/abs/1807.03749}
  {\bibfield  {journal} {\bibinfo  {journal} {arXiv [physics.chem-ph]}\ }
  (\bibinfo {year} {2018})},\ \Eprint {http://arxiv.org/abs/1807.03749}
  {arXiv:1807.03749} \BibitemShut {NoStop}%
\bibitem [{\citenamefont {Neufeld}\ and\ \citenamefont
  {Thom}(2017)}]{Neufeld2017b}%
  \BibitemOpen
  \bibfield  {author} {\bibinfo {author} {\bibfnamefont {V.~A.}\ \bibnamefont
  {Neufeld}}\ and\ \bibinfo {author} {\bibfnamefont {A.~J.~W.}\ \bibnamefont
  {Thom}},\ }\href {\doibase 10.1063/1.5003794} {\bibfield  {journal} {\bibinfo
   {journal} {J. Chem. Phys.}\ }\textbf {\bibinfo {volume} {147}},\ \bibinfo
  {pages} {194105} (\bibinfo {year} {2017})},\ \Eprint
  {http://arxiv.org/abs/1706.09923} {arXiv:1706.09923} \BibitemShut {NoStop}%
\bibitem [{\citenamefont {Booth}\ \emph {et~al.}(2009)\citenamefont {Booth},
  \citenamefont {Thom},\ and\ \citenamefont {Alavi}}]{Booth2009}%
  \BibitemOpen
  \bibfield  {author} {\bibinfo {author} {\bibfnamefont {G.~H.}\ \bibnamefont
  {Booth}}, \bibinfo {author} {\bibfnamefont {A.~J.~W.}\ \bibnamefont {Thom}},
  \ and\ \bibinfo {author} {\bibfnamefont {A.}~\bibnamefont {Alavi}},\ }\href
  {\doibase 10.1063/1.3193710} {\bibfield  {journal} {\bibinfo  {journal} {J.
  Chem. Phys.}\ }\textbf {\bibinfo {volume} {131}},\ \bibinfo {pages} {054106}
  (\bibinfo {year} {2009})}\BibitemShut {NoStop}%
\bibitem [{\citenamefont {Cleland}\ \emph {et~al.}(2010)\citenamefont
  {Cleland}, \citenamefont {Booth},\ and\ \citenamefont {Alavi}}]{Cleland2010}%
  \BibitemOpen
  \bibfield  {author} {\bibinfo {author} {\bibfnamefont {D.}~\bibnamefont
  {Cleland}}, \bibinfo {author} {\bibfnamefont {G.~H.}\ \bibnamefont {Booth}},
  \ and\ \bibinfo {author} {\bibfnamefont {A.}~\bibnamefont {Alavi}},\ }\href
  {\doibase 10.1063/1.3302277} {\bibfield  {journal} {\bibinfo  {journal} {J.
  Chem. Phys.}\ }\textbf {\bibinfo {volume} {132}},\ \bibinfo {pages} {041103}
  (\bibinfo {year} {2010})}\BibitemShut {NoStop}%
\bibitem [{\citenamefont {Spencer}\ \emph {et~al.}(2012)\citenamefont
  {Spencer}, \citenamefont {Blunt},\ and\ \citenamefont
  {Foulkes}}]{Spencer2012}%
  \BibitemOpen
  \bibfield  {author} {\bibinfo {author} {\bibfnamefont {J.~S.}\ \bibnamefont
  {Spencer}}, \bibinfo {author} {\bibfnamefont {N.~S.}\ \bibnamefont {Blunt}},
  \ and\ \bibinfo {author} {\bibfnamefont {W.~M.}\ \bibnamefont {Foulkes}},\
  }\href {\doibase 10.1063/1.3681396} {\bibfield  {journal} {\bibinfo
  {journal} {J. Chem. Phys.}\ }\textbf {\bibinfo {volume} {136}},\ \bibinfo
  {pages} {054110} (\bibinfo {year} {2012})},\ \Eprint
  {http://arxiv.org/abs/1110.5479} {arXiv:1110.5479} \BibitemShut {NoStop}%
\bibitem [{\citenamefont {Foulkes}\ \emph {et~al.}(2001)\citenamefont
  {Foulkes}, \citenamefont {Mitas}, \citenamefont {Needs},\ and\ \citenamefont
  {Rajagopal}}]{Foulkes2001}%
  \BibitemOpen
  \bibfield  {author} {\bibinfo {author} {\bibfnamefont {W.~M.~C.}\
  \bibnamefont {Foulkes}}, \bibinfo {author} {\bibfnamefont {L.}~\bibnamefont
  {Mitas}}, \bibinfo {author} {\bibfnamefont {R.~J.}\ \bibnamefont {Needs}}, \
  and\ \bibinfo {author} {\bibfnamefont {G.}~\bibnamefont {Rajagopal}},\ }\href
  {\doibase 10.1103/RevModPhys.73.33} {\bibfield  {journal} {\bibinfo
  {journal} {Rev. Mod. Phys.}\ }\textbf {\bibinfo {volume} {73}},\ \bibinfo
  {pages} {33} (\bibinfo {year} {2001})}\BibitemShut {NoStop}%
\bibitem [{\citenamefont {Booth}\ \emph {et~al.}(2011)\citenamefont {Booth},
  \citenamefont {Cleland}, \citenamefont {Thom},\ and\ \citenamefont
  {Alavi}}]{Booth2011}%
  \BibitemOpen
  \bibfield  {author} {\bibinfo {author} {\bibfnamefont {G.~H.}\ \bibnamefont
  {Booth}}, \bibinfo {author} {\bibfnamefont {D.}~\bibnamefont {Cleland}},
  \bibinfo {author} {\bibfnamefont {A.~J.~W.}\ \bibnamefont {Thom}}, \ and\
  \bibinfo {author} {\bibfnamefont {A.}~\bibnamefont {Alavi}},\ }\href
  {\doibase 10.1063/1.3624383} {\bibfield  {journal} {\bibinfo  {journal} {J.
  Chem. Phys.}\ }\textbf {\bibinfo {volume} {135}},\ \bibinfo {pages} {084104}
  (\bibinfo {year} {2011})}\BibitemShut {NoStop}%
\bibitem [{\citenamefont {Cleland}\ \emph {et~al.}(2012)\citenamefont
  {Cleland}, \citenamefont {Booth}, \citenamefont {Overy},\ and\ \citenamefont
  {Alavi}}]{Cleland2012}%
  \BibitemOpen
  \bibfield  {author} {\bibinfo {author} {\bibfnamefont {D.}~\bibnamefont
  {Cleland}}, \bibinfo {author} {\bibfnamefont {G.~H.}\ \bibnamefont {Booth}},
  \bibinfo {author} {\bibfnamefont {C.}~\bibnamefont {Overy}}, \ and\ \bibinfo
  {author} {\bibfnamefont {A.}~\bibnamefont {Alavi}},\ }\href {\doibase
  10.1021/ct300504f} {\bibfield  {journal} {\bibinfo  {journal} {J. Chem.
  Theory Comput.}\ }\textbf {\bibinfo {volume} {8}},\ \bibinfo {pages} {4138}
  (\bibinfo {year} {2012})}\BibitemShut {NoStop}%
\bibitem [{\citenamefont {Daday}\ \emph {et~al.}(2012)\citenamefont {Daday},
  \citenamefont {Smart}, \citenamefont {Booth}, \citenamefont {Alavi},\ and\
  \citenamefont {Filippi}}]{Daday2012}%
  \BibitemOpen
  \bibfield  {author} {\bibinfo {author} {\bibfnamefont {C.}~\bibnamefont
  {Daday}}, \bibinfo {author} {\bibfnamefont {S.}~\bibnamefont {Smart}},
  \bibinfo {author} {\bibfnamefont {G.~H.}\ \bibnamefont {Booth}}, \bibinfo
  {author} {\bibfnamefont {A.}~\bibnamefont {Alavi}}, \ and\ \bibinfo {author}
  {\bibfnamefont {C.}~\bibnamefont {Filippi}},\ }\href {\doibase
  10.1021/ct300486d} {\bibfield  {journal} {\bibinfo  {journal} {J. Chem.
  Theory Comput.}\ }\textbf {\bibinfo {volume} {8}},\ \bibinfo {pages} {4441}
  (\bibinfo {year} {2012})}\BibitemShut {NoStop}%
\bibitem [{\citenamefont {Booth}\ \emph {et~al.}(2014)\citenamefont {Booth},
  \citenamefont {Smart},\ and\ \citenamefont {Alavi}}]{Booth2014}%
  \BibitemOpen
  \bibfield  {author} {\bibinfo {author} {\bibfnamefont {G.~H.}\ \bibnamefont
  {Booth}}, \bibinfo {author} {\bibfnamefont {S.~D.}\ \bibnamefont {Smart}}, \
  and\ \bibinfo {author} {\bibfnamefont {A.}~\bibnamefont {Alavi}},\ }\href
  {\doibase 10.1080/00268976.2013.877165} {\bibfield  {journal} {\bibinfo
  {journal} {Mol. Phys.}\ }\textbf {\bibinfo {volume} {112}},\ \bibinfo {pages}
  {1855} (\bibinfo {year} {2014})}\BibitemShut {NoStop}%
\bibitem [{\citenamefont {Holmes}\ \emph
  {et~al.}(2016{\natexlab{a}})\citenamefont {Holmes}, \citenamefont
  {Changlani},\ and\ \citenamefont {Umrigar}}]{Holmes2016a}%
  \BibitemOpen
  \bibfield  {author} {\bibinfo {author} {\bibfnamefont {A.~A.}\ \bibnamefont
  {Holmes}}, \bibinfo {author} {\bibfnamefont {H.~J.}\ \bibnamefont
  {Changlani}}, \ and\ \bibinfo {author} {\bibfnamefont {C.~J.}\ \bibnamefont
  {Umrigar}},\ }\href {\doibase 10.1021/acs.jctc.5b01170} {\bibfield  {journal}
  {\bibinfo  {journal} {J. Chem. Theory Comput.}\ }\textbf {\bibinfo {volume}
  {12}},\ \bibinfo {pages} {1561} (\bibinfo {year}
  {2016}{\natexlab{a}})}\BibitemShut {NoStop}%
\bibitem [{\citenamefont {Sharma}\ \emph {et~al.}(2014)\citenamefont {Sharma},
  \citenamefont {Yanai}, \citenamefont {Booth}, \citenamefont {Umrigar},\ and\
  \citenamefont {Chan}}]{Sharma2014}%
  \BibitemOpen
  \bibfield  {author} {\bibinfo {author} {\bibfnamefont {S.}~\bibnamefont
  {Sharma}}, \bibinfo {author} {\bibfnamefont {T.}~\bibnamefont {Yanai}},
  \bibinfo {author} {\bibfnamefont {G.~H.}\ \bibnamefont {Booth}}, \bibinfo
  {author} {\bibfnamefont {C.~J.}\ \bibnamefont {Umrigar}}, \ and\ \bibinfo
  {author} {\bibfnamefont {G.~K.-L.}\ \bibnamefont {Chan}},\ }\href {\doibase
  10.1063/1.4867383} {\bibfield  {journal} {\bibinfo  {journal} {J. Chem.
  Phys.}\ }\textbf {\bibinfo {volume} {140}},\ \bibinfo {pages} {104112}
  (\bibinfo {year} {2014})}\BibitemShut {NoStop}%
\bibitem [{\citenamefont {Veis}\ \emph {et~al.}(2018)\citenamefont {Veis},
  \citenamefont {Antal{\'{i}}k}, \citenamefont {Legeza}, \citenamefont
  {Alavi},\ and\ \citenamefont {Pittner}}]{Veis2018}%
  \BibitemOpen
  \bibfield  {author} {\bibinfo {author} {\bibfnamefont {L.}~\bibnamefont
  {Veis}}, \bibinfo {author} {\bibfnamefont {A.}~\bibnamefont {Antal{\'{i}}k}},
  \bibinfo {author} {\bibfnamefont {{\"{O}}.}~\bibnamefont {Legeza}}, \bibinfo
  {author} {\bibfnamefont {A.}~\bibnamefont {Alavi}}, \ and\ \bibinfo {author}
  {\bibfnamefont {J.}~\bibnamefont {Pittner}},\ }\href
  {http://arxiv.org/abs/1801.01057} {\bibfield  {journal} {\bibinfo  {journal}
  {arXiv [physics.chem-ph]}\ } (\bibinfo {year} {2018})},\ \Eprint
  {http://arxiv.org/abs/1801.01057} {arXiv:1801.01057} \BibitemShut {NoStop}%
\bibitem [{\citenamefont {Samanta}\ \emph {et~al.}(2018)\citenamefont
  {Samanta}, \citenamefont {Blunt},\ and\ \citenamefont {Booth}}]{Samanta2018}%
  \BibitemOpen
  \bibfield  {author} {\bibinfo {author} {\bibfnamefont {P.~K.}\ \bibnamefont
  {Samanta}}, \bibinfo {author} {\bibfnamefont {N.~S.}\ \bibnamefont {Blunt}},
  \ and\ \bibinfo {author} {\bibfnamefont {G.~H.}\ \bibnamefont {Booth}},\
  }\href {\doibase 10.1021/acs.jctc.8b00454} {\bibfield  {journal} {\bibinfo
  {journal} {J. Chem. Theory Comput.}\ }\textbf {\bibinfo {volume} {14}},\
  \bibinfo {pages} {3532} (\bibinfo {year} {2018})}\BibitemShut {NoStop}%
\bibitem [{\citenamefont {Shepherd}\ \emph
  {et~al.}(2012{\natexlab{a}})\citenamefont {Shepherd}, \citenamefont
  {Gr{\"{u}}neis}, \citenamefont {Booth}, \citenamefont {Kresse},\ and\
  \citenamefont {Alavi}}]{Shepherd2012}%
  \BibitemOpen
  \bibfield  {author} {\bibinfo {author} {\bibfnamefont {J.~J.}\ \bibnamefont
  {Shepherd}}, \bibinfo {author} {\bibfnamefont {A.}~\bibnamefont
  {Gr{\"{u}}neis}}, \bibinfo {author} {\bibfnamefont {G.~H.}\ \bibnamefont
  {Booth}}, \bibinfo {author} {\bibfnamefont {G.}~\bibnamefont {Kresse}}, \
  and\ \bibinfo {author} {\bibfnamefont {A.}~\bibnamefont {Alavi}},\ }\href
  {\doibase 10.1103/PhysRevB.86.035111} {\bibfield  {journal} {\bibinfo
  {journal} {Phys. Rev. B}\ }\textbf {\bibinfo {volume} {86}},\ \bibinfo
  {pages} {035111} (\bibinfo {year} {2012}{\natexlab{a}})},\ \Eprint
  {http://arxiv.org/abs/1202.4990} {arXiv:1202.4990} \BibitemShut {NoStop}%
\bibitem [{\citenamefont {Shepherd}\ \emph
  {et~al.}(2012{\natexlab{b}})\citenamefont {Shepherd}, \citenamefont {Booth},
  \citenamefont {Gr{\"{u}}neis},\ and\ \citenamefont {Alavi}}]{Shepherd2012a}%
  \BibitemOpen
  \bibfield  {author} {\bibinfo {author} {\bibfnamefont {J.~J.}\ \bibnamefont
  {Shepherd}}, \bibinfo {author} {\bibfnamefont {G.}~\bibnamefont {Booth}},
  \bibinfo {author} {\bibfnamefont {A.}~\bibnamefont {Gr{\"{u}}neis}}, \ and\
  \bibinfo {author} {\bibfnamefont {A.}~\bibnamefont {Alavi}},\ }\href
  {\doibase 10.1103/PhysRevB.85.081103} {\bibfield  {journal} {\bibinfo
  {journal} {Phys. Rev. B}\ }\textbf {\bibinfo {volume} {85}},\ \bibinfo
  {pages} {081103} (\bibinfo {year} {2012}{\natexlab{b}})},\ \Eprint
  {http://arxiv.org/abs/1109.2635} {arXiv:1109.2635} \BibitemShut {NoStop}%
\bibitem [{\citenamefont {Shepherd}\ \emph
  {et~al.}(2012{\natexlab{c}})\citenamefont {Shepherd}, \citenamefont {Booth},\
  and\ \citenamefont {Alavi}}]{Shepherd2012b}%
  \BibitemOpen
  \bibfield  {author} {\bibinfo {author} {\bibfnamefont {J.~J.}\ \bibnamefont
  {Shepherd}}, \bibinfo {author} {\bibfnamefont {G.~H.}\ \bibnamefont {Booth}},
  \ and\ \bibinfo {author} {\bibfnamefont {A.}~\bibnamefont {Alavi}},\ }\href
  {\doibase 10.1063/1.4720076} {\bibfield  {journal} {\bibinfo  {journal} {J.
  Chem. Phys.}\ }\textbf {\bibinfo {volume} {136}},\ \bibinfo {pages} {244101}
  (\bibinfo {year} {2012}{\natexlab{c}})},\ \Eprint
  {http://arxiv.org/abs/1201.4691} {arXiv:1201.4691} \BibitemShut {NoStop}%
\bibitem [{\citenamefont {Booth}\ \emph {et~al.}(2013)\citenamefont {Booth},
  \citenamefont {Gr{\"{u}}neis}, \citenamefont {Kresse},\ and\ \citenamefont
  {Alavi}}]{Booth2013}%
  \BibitemOpen
  \bibfield  {author} {\bibinfo {author} {\bibfnamefont {G.~H.}\ \bibnamefont
  {Booth}}, \bibinfo {author} {\bibfnamefont {A.}~\bibnamefont
  {Gr{\"{u}}neis}}, \bibinfo {author} {\bibfnamefont {G.}~\bibnamefont
  {Kresse}}, \ and\ \bibinfo {author} {\bibfnamefont {A.}~\bibnamefont
  {Alavi}},\ }\href {\doibase 10.1038/nature11770} {\bibfield  {journal}
  {\bibinfo  {journal} {Nature}\ }\textbf {\bibinfo {volume} {493}},\ \bibinfo
  {pages} {365} (\bibinfo {year} {2013})}\BibitemShut {NoStop}%
\bibitem [{\citenamefont {Schwarz}\ \emph {et~al.}(2015)\citenamefont
  {Schwarz}, \citenamefont {Booth},\ and\ \citenamefont {Alavi}}]{Schwarz2015}%
  \BibitemOpen
  \bibfield  {author} {\bibinfo {author} {\bibfnamefont {L.~R.}\ \bibnamefont
  {Schwarz}}, \bibinfo {author} {\bibfnamefont {G.~H.}\ \bibnamefont {Booth}},
  \ and\ \bibinfo {author} {\bibfnamefont {A.}~\bibnamefont {Alavi}},\ }\href
  {\doibase 10.1103/PhysRevB.91.045139} {\bibfield  {journal} {\bibinfo
  {journal} {Phys. Rev. B}\ }\textbf {\bibinfo {volume} {91}},\ \bibinfo
  {pages} {045139} (\bibinfo {year} {2015})}\BibitemShut {NoStop}%
\bibitem [{\citenamefont {Luo}\ and\ \citenamefont {Alavi}(2018)}]{Luo2018a}%
  \BibitemOpen
  \bibfield  {author} {\bibinfo {author} {\bibfnamefont {H.}~\bibnamefont
  {Luo}}\ and\ \bibinfo {author} {\bibfnamefont {A.}~\bibnamefont {Alavi}},\
  }\href {\doibase 10.1021/acs.jctc.7b01257} {\bibfield  {journal} {\bibinfo
  {journal} {J. Chem. Theory Comput.}\ }\textbf {\bibinfo {volume} {14}},\
  \bibinfo {pages} {1403} (\bibinfo {year} {2018})}\BibitemShut {NoStop}%
\bibitem [{\citenamefont {Ruggeri}\ \emph {et~al.}(2018)\citenamefont
  {Ruggeri}, \citenamefont {R{\'{i}}os},\ and\ \citenamefont
  {Alavi}}]{Ruggeri2018}%
  \BibitemOpen
  \bibfield  {author} {\bibinfo {author} {\bibfnamefont {M.}~\bibnamefont
  {Ruggeri}}, \bibinfo {author} {\bibfnamefont {P.~L.}\ \bibnamefont
  {R{\'{i}}os}}, \ and\ \bibinfo {author} {\bibfnamefont {A.}~\bibnamefont
  {Alavi}},\ }\href {https://arxiv.org/pdf/1804.04938.pdf
  http://arxiv.org/abs/1804.04938} {\bibfield  {journal} {\bibinfo  {journal}
  {arXiv[cond-mat.str-el]}\ } (\bibinfo {year} {2018})},\ \Eprint
  {http://arxiv.org/abs/1804.04938} {arXiv:1804.04938} \BibitemShut {NoStop}%
\bibitem [{\citenamefont {Booth}\ and\ \citenamefont {Chan}(2012)}]{Booth2012}%
  \BibitemOpen
  \bibfield  {author} {\bibinfo {author} {\bibfnamefont {G.~H.}\ \bibnamefont
  {Booth}}\ and\ \bibinfo {author} {\bibfnamefont {G.~K.-L.}\ \bibnamefont
  {Chan}},\ }\href {\doibase 10.1063/1.4766327} {\bibfield  {journal} {\bibinfo
   {journal} {J. Chem. Phys.}\ }\textbf {\bibinfo {volume} {137}},\ \bibinfo
  {pages} {191102} (\bibinfo {year} {2012})}\BibitemShut {NoStop}%
\bibitem [{\citenamefont {Ten-no}(2013)}]{Ten-no2013}%
  \BibitemOpen
  \bibfield  {author} {\bibinfo {author} {\bibfnamefont {S.}~\bibnamefont
  {Ten-no}},\ }\href {\doibase 10.1063/1.4802766} {\bibfield  {journal}
  {\bibinfo  {journal} {J. Chem. Phys.}\ }\textbf {\bibinfo {volume} {138}},\
  \bibinfo {pages} {164126} (\bibinfo {year} {2013})}\BibitemShut {NoStop}%
\bibitem [{\citenamefont {Humeniuk}\ and\ \citenamefont
  {Mitri{\'{c}}}(2014)}]{Humeniuk2014}%
  \BibitemOpen
  \bibfield  {author} {\bibinfo {author} {\bibfnamefont {A.}~\bibnamefont
  {Humeniuk}}\ and\ \bibinfo {author} {\bibfnamefont {R.}~\bibnamefont
  {Mitri{\'{c}}}},\ }\href {\doibase 10.1063/1.4901020} {\bibfield  {journal}
  {\bibinfo  {journal} {J. Chem. Phys.}\ }\textbf {\bibinfo {volume} {141}},\
  \bibinfo {pages} {194104} (\bibinfo {year} {2014})}\BibitemShut {NoStop}%
\bibitem [{\citenamefont {Blunt}\ \emph
  {et~al.}(2015{\natexlab{a}})\citenamefont {Blunt}, \citenamefont {Smart},
  \citenamefont {Booth},\ and\ \citenamefont {Alavi}}]{Blunt2015}%
  \BibitemOpen
  \bibfield  {author} {\bibinfo {author} {\bibfnamefont {N.~S.}\ \bibnamefont
  {Blunt}}, \bibinfo {author} {\bibfnamefont {S.~D.}\ \bibnamefont {Smart}},
  \bibinfo {author} {\bibfnamefont {G.~H.}\ \bibnamefont {Booth}}, \ and\
  \bibinfo {author} {\bibfnamefont {A.}~\bibnamefont {Alavi}},\ }\href
  {\doibase 10.1063/1.4932595} {\bibfield  {journal} {\bibinfo  {journal} {J.
  Chem. Phys.}\ }\textbf {\bibinfo {volume} {143}},\ \bibinfo {pages} {134117}
  (\bibinfo {year} {2015}{\natexlab{a}})}\BibitemShut {NoStop}%
\bibitem [{\citenamefont {Blunt}\ \emph {et~al.}(2017)\citenamefont {Blunt},
  \citenamefont {Booth},\ and\ \citenamefont {Alavi}}]{Blunt2017}%
  \BibitemOpen
  \bibfield  {author} {\bibinfo {author} {\bibfnamefont {N.~S.}\ \bibnamefont
  {Blunt}}, \bibinfo {author} {\bibfnamefont {G.~H.}\ \bibnamefont {Booth}}, \
  and\ \bibinfo {author} {\bibfnamefont {A.}~\bibnamefont {Alavi}},\ }\href
  {\doibase 10.1063/1.4986963} {\bibfield  {journal} {\bibinfo  {journal} {J.
  Chem. Phys.}\ }\textbf {\bibinfo {volume} {146}},\ \bibinfo {pages} {244105}
  (\bibinfo {year} {2017})}\BibitemShut {NoStop}%
\bibitem [{\citenamefont {Deustua}\ \emph {et~al.}(2017)\citenamefont
  {Deustua}, \citenamefont {Shen},\ and\ \citenamefont
  {Piecuch}}]{Deustua2017}%
  \BibitemOpen
  \bibfield  {author} {\bibinfo {author} {\bibfnamefont {J.~E.}\ \bibnamefont
  {Deustua}}, \bibinfo {author} {\bibfnamefont {J.}~\bibnamefont {Shen}}, \
  and\ \bibinfo {author} {\bibfnamefont {P.}~\bibnamefont {Piecuch}},\ }\href
  {\doibase 10.1103/PhysRevLett.119.223003} {\bibfield  {journal} {\bibinfo
  {journal} {Phys. Rev. Lett.}\ }\textbf {\bibinfo {volume} {119}},\ \bibinfo
  {pages} {223003} (\bibinfo {year} {2017})}\BibitemShut {NoStop}%
\bibitem [{\citenamefont {Petruzielo}\ \emph {et~al.}(2012)\citenamefont
  {Petruzielo}, \citenamefont {Holmes}, \citenamefont {Changlani},
  \citenamefont {Nightingale},\ and\ \citenamefont {Umrigar}}]{Petruzielo2012}%
  \BibitemOpen
  \bibfield  {author} {\bibinfo {author} {\bibfnamefont {F.~R.}\ \bibnamefont
  {Petruzielo}}, \bibinfo {author} {\bibfnamefont {A.~A.}\ \bibnamefont
  {Holmes}}, \bibinfo {author} {\bibfnamefont {H.~J.}\ \bibnamefont
  {Changlani}}, \bibinfo {author} {\bibfnamefont {M.~P.}\ \bibnamefont
  {Nightingale}}, \ and\ \bibinfo {author} {\bibfnamefont {C.~J.}\ \bibnamefont
  {Umrigar}},\ }\href {\doibase 10.1103/PhysRevLett.109.230201} {\bibfield
  {journal} {\bibinfo  {journal} {Phys. Rev. Lett.}\ }\textbf {\bibinfo
  {volume} {109}},\ \bibinfo {pages} {230201} (\bibinfo {year}
  {2012})}\BibitemShut {NoStop}%
\bibitem [{\citenamefont {Blunt}\ \emph
  {et~al.}(2015{\natexlab{b}})\citenamefont {Blunt}, \citenamefont {Smart},
  \citenamefont {Kersten}, \citenamefont {Spencer}, \citenamefont {Booth},\
  and\ \citenamefont {Alavi}}]{Blunt2015b}%
  \BibitemOpen
  \bibfield  {author} {\bibinfo {author} {\bibfnamefont {N.~S.}\ \bibnamefont
  {Blunt}}, \bibinfo {author} {\bibfnamefont {S.~D.}\ \bibnamefont {Smart}},
  \bibinfo {author} {\bibfnamefont {J.~A.~F.}\ \bibnamefont {Kersten}},
  \bibinfo {author} {\bibfnamefont {J.~S.}\ \bibnamefont {Spencer}}, \bibinfo
  {author} {\bibfnamefont {G.~H.}\ \bibnamefont {Booth}}, \ and\ \bibinfo
  {author} {\bibfnamefont {A.}~\bibnamefont {Alavi}},\ }\href {\doibase
  10.1063/1.4920975} {\bibfield  {journal} {\bibinfo  {journal} {J. Chem.
  Phys.}\ }\textbf {\bibinfo {volume} {142}},\ \bibinfo {pages} {184107}
  (\bibinfo {year} {2015}{\natexlab{b}})}\BibitemShut {NoStop}%
\bibitem [{\citenamefont {Kersten}\ \emph {et~al.}(2016)\citenamefont
  {Kersten}, \citenamefont {Booth},\ and\ \citenamefont {Alavi}}]{Kersten2016}%
  \BibitemOpen
  \bibfield  {author} {\bibinfo {author} {\bibfnamefont {J.~A.~F.}\
  \bibnamefont {Kersten}}, \bibinfo {author} {\bibfnamefont {G.~H.}\
  \bibnamefont {Booth}}, \ and\ \bibinfo {author} {\bibfnamefont
  {A.}~\bibnamefont {Alavi}},\ }\href {\doibase 10.1063/1.4959245} {\bibfield
  {journal} {\bibinfo  {journal} {J. Chem. Phys.}\ }\textbf {\bibinfo {volume}
  {145}},\ \bibinfo {pages} {054117} (\bibinfo {year} {2016})}\BibitemShut
  {NoStop}%
\bibitem [{\citenamefont {Smart}\ \emph {et~al.}()\citenamefont {Smart},
  \citenamefont {Booth},\ and\ \citenamefont {Alavi}}]{Smartunpub}%
  \BibitemOpen
  \bibfield  {author} {\bibinfo {author} {\bibfnamefont {S.~D.}\ \bibnamefont
  {Smart}}, \bibinfo {author} {\bibfnamefont {G.~H.}\ \bibnamefont {Booth}}, \
  and\ \bibinfo {author} {\bibfnamefont {A.}~\bibnamefont {Alavi}},\
  }\href@noop {} {\bibinfo  {journal} {unpublished}\ }\BibitemShut {NoStop}%
\bibitem [{\citenamefont {Blunt}(2018)}]{Blunt2018}%
  \BibitemOpen
\bibfield  {journal} {  }\bibfield  {author} {\bibinfo {author} {\bibfnamefont
  {N.~S.}\ \bibnamefont {Blunt}},\ }\href {\doibase 10.1063/1.5037923}
  {\bibfield  {journal} {\bibinfo  {journal} {J. Chem. Phys.}\ }\textbf
  {\bibinfo {volume} {148}},\ \bibinfo {pages} {221101} (\bibinfo {year}
  {2018})}\BibitemShut {NoStop}%
\bibitem [{\citenamefont {Holmes}\ \emph
  {et~al.}(2016{\natexlab{b}})\citenamefont {Holmes}, \citenamefont {Tubman},\
  and\ \citenamefont {Umrigar}}]{Holmes2016}%
  \BibitemOpen
  \bibfield  {author} {\bibinfo {author} {\bibfnamefont {A.~A.}\ \bibnamefont
  {Holmes}}, \bibinfo {author} {\bibfnamefont {N.~M.}\ \bibnamefont {Tubman}},
  \ and\ \bibinfo {author} {\bibfnamefont {C.~J.}\ \bibnamefont {Umrigar}},\
  }\href {\doibase 10.1021/acs.jctc.6b00407} {\bibfield  {journal} {\bibinfo
  {journal} {J. Chem. Theory Comput.}\ }\textbf {\bibinfo {volume} {12}},\
  \bibinfo {pages} {3674} (\bibinfo {year} {2016}{\natexlab{b}})},\ \Eprint
  {http://arxiv.org/abs/1606.07453} {arXiv:1606.07453} \BibitemShut {NoStop}%
\bibitem [{\citenamefont {Sharma}\ \emph {et~al.}(2017)\citenamefont {Sharma},
  \citenamefont {Holmes}, \citenamefont {Jeanmairet}, \citenamefont {Alavi},\
  and\ \citenamefont {Umrigar}}]{Sharma2017}%
  \BibitemOpen
  \bibfield  {author} {\bibinfo {author} {\bibfnamefont {S.}~\bibnamefont
  {Sharma}}, \bibinfo {author} {\bibfnamefont {A.~A.}\ \bibnamefont {Holmes}},
  \bibinfo {author} {\bibfnamefont {G.}~\bibnamefont {Jeanmairet}}, \bibinfo
  {author} {\bibfnamefont {A.}~\bibnamefont {Alavi}}, \ and\ \bibinfo {author}
  {\bibfnamefont {C.~J.}\ \bibnamefont {Umrigar}},\ }\href {\doibase
  10.1021/acs.jctc.6b01028} {\bibfield  {journal} {\bibinfo  {journal} {J.
  Chem. Theory Comput.}\ }\textbf {\bibinfo {volume} {13}},\ \bibinfo {pages}
  {1595} (\bibinfo {year} {2017})}\BibitemShut {NoStop}%
\bibitem [{\citenamefont {Holmes}\ \emph {et~al.}(2017)\citenamefont {Holmes},
  \citenamefont {Umrigar},\ and\ \citenamefont {Sharma}}]{Holmes2017}%
  \BibitemOpen
  \bibfield  {author} {\bibinfo {author} {\bibfnamefont {A.~A.}\ \bibnamefont
  {Holmes}}, \bibinfo {author} {\bibfnamefont {C.~J.}\ \bibnamefont {Umrigar}},
  \ and\ \bibinfo {author} {\bibfnamefont {S.}~\bibnamefont {Sharma}},\ }\href
  {\doibase 10.1063/1.4998614} {\bibfield  {journal} {\bibinfo  {journal} {J.
  Chem. Phys.}\ }\textbf {\bibinfo {volume} {147}},\ \bibinfo {pages} {164111}
  (\bibinfo {year} {2017})}\BibitemShut {NoStop}%
\bibitem [{\citenamefont {Power}\ and\ \citenamefont
  {Pitzer}(1974)}]{Power1974}%
  \BibitemOpen
  \bibfield  {author} {\bibinfo {author} {\bibfnamefont {J.~D.}\ \bibnamefont
  {Power}}\ and\ \bibinfo {author} {\bibfnamefont {R.~M.}\ \bibnamefont
  {Pitzer}},\ }\href
  {http://ac.els-cdn.com/0009261474801594/1-s2.0-0009261474801594-main.pdf?{\_}tid=cccf3aae-203f-11e7-b941-00000aab0f26{\&}acdnat=1492084585{\_}3a6648ebc0180a56b03a630fae748a77}
  {\bibfield  {journal} {\bibinfo  {journal} {Chem. Phys. Lett.}\ }\textbf
  {\bibinfo {volume} {24}},\ \bibinfo {pages} {478} (\bibinfo {year}
  {1974})}\BibitemShut {NoStop}%
\bibitem [{\citenamefont {Helgaker}\ \emph {et~al.}(2014)\citenamefont
  {Helgaker}, \citenamefont {J{\o}rgensen},\ and\ \citenamefont
  {Olsen}}]{Helgaker2000}%
  \BibitemOpen
  \bibfield  {author} {\bibinfo {author} {\bibfnamefont {T.}~\bibnamefont
  {Helgaker}}, \bibinfo {author} {\bibfnamefont {P.}~\bibnamefont
  {J{\o}rgensen}}, \ and\ \bibinfo {author} {\bibfnamefont {J.}~\bibnamefont
  {Olsen}},\ }in\ \href {\doibase 10.1002/9781119019572.ch13} {\emph {\bibinfo
  {booktitle} {Mol. Electron. Theory}}}\ (\bibinfo  {publisher} {John Wiley
  {\&} Sons, Ltd},\ \bibinfo {address} {Chichester, UK},\ \bibinfo {year}
  {2014})\ pp.\ \bibinfo {pages} {648--723}\BibitemShut {NoStop}%
\bibitem [{\citenamefont {Overy}\ \emph {et~al.}(2014)\citenamefont {Overy},
  \citenamefont {Booth}, \citenamefont {Blunt}, \citenamefont {Shepherd},
  \citenamefont {Cleland},\ and\ \citenamefont {Alavi}}]{Overy2014}%
  \BibitemOpen
  \bibfield  {author} {\bibinfo {author} {\bibfnamefont {C.}~\bibnamefont
  {Overy}}, \bibinfo {author} {\bibfnamefont {G.~H.}\ \bibnamefont {Booth}},
  \bibinfo {author} {\bibfnamefont {N.~S.}\ \bibnamefont {Blunt}}, \bibinfo
  {author} {\bibfnamefont {J.~J.}\ \bibnamefont {Shepherd}}, \bibinfo {author}
  {\bibfnamefont {D.}~\bibnamefont {Cleland}}, \ and\ \bibinfo {author}
  {\bibfnamefont {A.}~\bibnamefont {Alavi}},\ }\href {\doibase
  10.1063/1.4904313} {\bibfield  {journal} {\bibinfo  {journal} {J. Chem.
  Phys.}\ }\textbf {\bibinfo {volume} {141}},\ \bibinfo {pages} {244117}
  (\bibinfo {year} {2014})}\BibitemShut {NoStop}%
\bibitem [{\citenamefont {Umrigar}\ \emph {et~al.}(1993)\citenamefont
  {Umrigar}, \citenamefont {Nightingale},\ and\ \citenamefont
  {Runge}}]{Umrigar1993}%
  \BibitemOpen
  \bibfield  {author} {\bibinfo {author} {\bibfnamefont {C.~J.}\ \bibnamefont
  {Umrigar}}, \bibinfo {author} {\bibfnamefont {M.~P.}\ \bibnamefont
  {Nightingale}}, \ and\ \bibinfo {author} {\bibfnamefont {K.~J.}\ \bibnamefont
  {Runge}},\ }\href {\doibase 10.1063/1.465195} {\bibfield  {journal} {\bibinfo
   {journal} {J. Chem. Phys.}\ }\textbf {\bibinfo {volume} {99}},\ \bibinfo
  {pages} {2865} (\bibinfo {year} {1993})}\BibitemShut {NoStop}%
\bibitem [{\citenamefont {Vigor}\ \emph {et~al.}(2015)\citenamefont {Vigor},
  \citenamefont {Spencer}, \citenamefont {Bearpark},\ and\ \citenamefont
  {Thom}}]{Vigor2015}%
  \BibitemOpen
  \bibfield  {author} {\bibinfo {author} {\bibfnamefont {W.~A.}\ \bibnamefont
  {Vigor}}, \bibinfo {author} {\bibfnamefont {J.~S.}\ \bibnamefont {Spencer}},
  \bibinfo {author} {\bibfnamefont {M.~J.}\ \bibnamefont {Bearpark}}, \ and\
  \bibinfo {author} {\bibfnamefont {A.~J.~W.}\ \bibnamefont {Thom}},\ }\href
  {\doibase 10.1063/1.4913644} {\bibfield  {journal} {\bibinfo  {journal} {J.
  Chem. Phys.}\ }\textbf {\bibinfo {volume} {142}},\ \bibinfo {pages} {104101}
  (\bibinfo {year} {2015})},\ \Eprint {http://arxiv.org/abs/1407.1753}
  {arXiv:1407.1753} \BibitemShut {NoStop}%
\bibitem [{\citenamefont {Flyvbjerg}\ and\ \citenamefont
  {Petersen}(1989)}]{Flyvbjerg1989}%
  \BibitemOpen
  \bibfield  {author} {\bibinfo {author} {\bibfnamefont {H.}~\bibnamefont
  {Flyvbjerg}}\ and\ \bibinfo {author} {\bibfnamefont {H.~G.}\ \bibnamefont
  {Petersen}},\ }\href {\doibase 10.1063/1.457480} {\bibfield  {journal}
  {\bibinfo  {journal} {J. Chem. Phys.}\ }\textbf {\bibinfo {volume} {91}},\
  \bibinfo {pages} {461} (\bibinfo {year} {1989})}\BibitemShut {NoStop}%
\bibitem [{Note1()}]{Note1}%
  \BibitemOpen
  \bibinfo {note} {For code, see \protect \url
  {https://github.com/jsspencer/pyblock}}\BibitemShut {NoStop}%
\bibitem [{\citenamefont {Spencer}\ \emph {et~al.}(2015)\citenamefont
  {Spencer}, \citenamefont {Blunt}, \citenamefont {Vigor}, \citenamefont
  {Malone}, \citenamefont {Foulkes}, \citenamefont {Shepherd},\ and\
  \citenamefont {Thom}}]{HANDEpaper}%
  \BibitemOpen
  \bibfield  {author} {\bibinfo {author} {\bibfnamefont {J.~S.}\ \bibnamefont
  {Spencer}}, \bibinfo {author} {\bibfnamefont {N.~S.}\ \bibnamefont {Blunt}},
  \bibinfo {author} {\bibfnamefont {W.~A.}\ \bibnamefont {Vigor}}, \bibinfo
  {author} {\bibfnamefont {F.~D.}\ \bibnamefont {Malone}}, \bibinfo {author}
  {\bibfnamefont {W.~M.~C.}\ \bibnamefont {Foulkes}}, \bibinfo {author}
  {\bibfnamefont {J.~J.}\ \bibnamefont {Shepherd}}, \ and\ \bibinfo {author}
  {\bibfnamefont {A.~J.~W.}\ \bibnamefont {Thom}},\ }\href {\doibase
  10.5334/jors.bw} {\bibfield  {journal} {\bibinfo  {journal} {J. Open Res.
  Softw.}\ }\textbf {\bibinfo {volume} {3}},\ \bibinfo {pages} {1} (\bibinfo
  {year} {2015})},\ \Eprint {http://arxiv.org/abs/1407.5407} {arXiv:1407.5407}
  \BibitemShut {NoStop}%
\bibitem [{Note2()}]{Note2}%
  \BibitemOpen
  \bibinfo {note} {See \protect \url {http://www.hande.org.uk/} and \protect
  \url {https://github.com/hande-qmc/hande} for information and
  code}\BibitemShut {NoStop}%
\bibitem [{\citenamefont {Thom}\ and\ \citenamefont {Alavi}(2005)}]{Thom2005}%
  \BibitemOpen
  \bibfield  {author} {\bibinfo {author} {\bibfnamefont {A.~J.~W.}\
  \bibnamefont {Thom}}\ and\ \bibinfo {author} {\bibfnamefont {A.}~\bibnamefont
  {Alavi}},\ }\href {\doibase 10.1063/1.2114849} {\bibfield  {journal}
  {\bibinfo  {journal} {J. Chem. Phys.}\ }\textbf {\bibinfo {volume} {123}},\
  \bibinfo {pages} {204106} (\bibinfo {year} {2005})}\BibitemShut {NoStop}%
\bibitem [{\citenamefont {Thom}\ and\ \citenamefont {Alavi}(2007)}]{Thom2007}%
  \BibitemOpen
  \bibfield  {author} {\bibinfo {author} {\bibfnamefont {A.~J.~W.}\
  \bibnamefont {Thom}}\ and\ \bibinfo {author} {\bibfnamefont {A.}~\bibnamefont
  {Alavi}},\ }\href {\doibase 10.1103/PhysRevLett.99.143001} {\bibfield
  {journal} {\bibinfo  {journal} {Phys. Rev. Lett.}\ }\textbf {\bibinfo
  {volume} {99}},\ \bibinfo {pages} {143001} (\bibinfo {year}
  {2007})}\BibitemShut {NoStop}%
\bibitem [{\citenamefont {Kolodrubetz}\ and\ \citenamefont
  {Clark}(2012)}]{Kolodrubetz2012}%
  \BibitemOpen
  \bibfield  {author} {\bibinfo {author} {\bibfnamefont {M.}~\bibnamefont
  {Kolodrubetz}}\ and\ \bibinfo {author} {\bibfnamefont {B.~K.}\ \bibnamefont
  {Clark}},\ }\href {\doibase 10.1103/PhysRevB.86.075109} {\bibfield  {journal}
  {\bibinfo  {journal} {Phys. Rev. B}\ }\textbf {\bibinfo {volume} {86}},\
  \bibinfo {pages} {075109} (\bibinfo {year} {2012})}\BibitemShut {NoStop}%
\bibitem [{\citenamefont {Pederiva}\ \emph {et~al.}(2017)\citenamefont
  {Pederiva}, \citenamefont {Roggero},\ and\ \citenamefont
  {Schmidt}}]{Pederiva2017}%
  \BibitemOpen
  \bibfield  {author} {\bibinfo {author} {\bibfnamefont {F.}~\bibnamefont
  {Pederiva}}, \bibinfo {author} {\bibfnamefont {A.}~\bibnamefont {Roggero}}, \
  and\ \bibinfo {author} {\bibfnamefont {K.~E.}\ \bibnamefont {Schmidt}},\
  }\enquote {\bibinfo {title} {{Variational and Diffusion Monte Carlo
  Approaches to the Nuclear Few- and Many-Body Problem}},}\ in\ \href {\doibase
  10.1007/978-3-319-53336-0_9} {\emph {\bibinfo {booktitle} {An Adv. Course
  Comput. Nucl. Phys. Bridg. Scales from Quarks to Neutron Stars}}},\ \bibinfo
  {editor} {edited by\ \bibinfo {editor} {\bibfnamefont {M.}~\bibnamefont
  {Hjorth-Jensen}}, \bibinfo {editor} {\bibfnamefont {M.~P.}\ \bibnamefont
  {Lombardo}}, \ and\ \bibinfo {editor} {\bibfnamefont {U.}~\bibnamefont {van
  Kolck}}}\ (\bibinfo  {publisher} {Springer International Publishing},\
  \bibinfo {address} {Cham},\ \bibinfo {year} {2017})\ pp.\ \bibinfo {pages}
  {401--476}\BibitemShut {NoStop}%
\bibitem [{\citenamefont {Ohtsuka}\ and\ \citenamefont
  {Nagase}(2008)}]{Ohtsuka2008}%
  \BibitemOpen
  \bibfield  {author} {\bibinfo {author} {\bibfnamefont {Y.}~\bibnamefont
  {Ohtsuka}}\ and\ \bibinfo {author} {\bibfnamefont {S.}~\bibnamefont
  {Nagase}},\ }\href {\doibase 10.1016/j.cplett.2008.08.090} {\bibfield
  {journal} {\bibinfo  {journal} {Chem. Phys. Lett.}\ }\textbf {\bibinfo
  {volume} {463}},\ \bibinfo {pages} {431} (\bibinfo {year}
  {2008})}\BibitemShut {NoStop}%
\bibitem [{Note3()}]{Note3}%
  \BibitemOpen
  \bibinfo {note} {Idea by Alavi and co--workers, this was suggested to us as
  an alternative by Pablo L\'opez R\'ios (personal communication).}\BibitemShut
  {Stop}%
\bibitem [{\citenamefont {Walker}(1974)}]{Walker1974}%
  \BibitemOpen
  \bibfield  {author} {\bibinfo {author} {\bibfnamefont {A.~J.}\ \bibnamefont
  {Walker}},\ }\href {\doibase 10.1049/el:19740097} {\bibfield  {journal}
  {\bibinfo  {journal} {Electron. Lett.}\ }\textbf {\bibinfo {volume} {10}},\
  \bibinfo {pages} {127} (\bibinfo {year} {1974})}\BibitemShut {NoStop}%
\bibitem [{\citenamefont {Walker}(1977)}]{Walker1977}%
  \BibitemOpen
  \bibfield  {author} {\bibinfo {author} {\bibfnamefont {A.~J.}\ \bibnamefont
  {Walker}},\ }\href {\doibase 10.1145/355744.355749} {\bibfield  {journal}
  {\bibinfo  {journal} {ACM Trans. Math. Softw.}\ }\textbf {\bibinfo {volume}
  {3}},\ \bibinfo {pages} {253} (\bibinfo {year} {1977})}\BibitemShut {NoStop}%
\bibitem [{\citenamefont {Kronmal}\ and\ \citenamefont
  {Peterson}(1979)}]{Kronmal1979}%
  \BibitemOpen
  \bibfield  {author} {\bibinfo {author} {\bibfnamefont {R.~A.}\ \bibnamefont
  {Kronmal}}\ and\ \bibinfo {author} {\bibfnamefont {A.~V.}\ \bibnamefont
  {Peterson}},\ }\href {\doibase 10.1080/00031305.1979.10482697} {\bibfield
  {journal} {\bibinfo  {journal} {Am. Stat.}\ }\textbf {\bibinfo {volume}
  {33}},\ \bibinfo {pages} {214} (\bibinfo {year} {1979})}\BibitemShut
  {NoStop}%
\bibitem [{Note4()}]{Note4}%
  \BibitemOpen
  \bibinfo {note} {It is not clear from Holmes et al. \cite {Holmes2016a} what
  happens if $H_{ia} = H_{ija}$}\BibitemShut {NoStop}%
\bibitem [{Note5()}]{Note5}%
  \BibitemOpen
  \bibinfo {note} {Personal communication with Ali Alavi and Pablo L\'opez
  R\'ios.}\BibitemShut {Stop}%
\bibitem [{\citenamefont {Slater}(1929)}]{Slater1929}%
  \BibitemOpen
  \bibfield  {author} {\bibinfo {author} {\bibfnamefont {J.~C.}\ \bibnamefont
  {Slater}},\ }\href {\doibase 10.1103/PhysRev.34.1293} {\bibfield  {journal}
  {\bibinfo  {journal} {Phys. Rev.}\ }\textbf {\bibinfo {volume} {34}},\
  \bibinfo {pages} {1293} (\bibinfo {year} {1929})}\BibitemShut {NoStop}%
\bibitem [{\citenamefont {Condon}(1930)}]{Condon1930}%
  \BibitemOpen
  \bibfield  {author} {\bibinfo {author} {\bibfnamefont {E.~U.}\ \bibnamefont
  {Condon}},\ }\href {\doibase 10.1103/PhysRev.36.1121} {\bibfield  {journal}
  {\bibinfo  {journal} {Phys. Rev.}\ }\textbf {\bibinfo {volume} {36}},\
  \bibinfo {pages} {1121} (\bibinfo {year} {1930})}\BibitemShut {NoStop}%
\bibitem [{\citenamefont {Whitten}(1973)}]{Whitten1973}%
  \BibitemOpen
  \bibfield  {author} {\bibinfo {author} {\bibfnamefont {J.~L.}\ \bibnamefont
  {Whitten}},\ }\href {\doibase 10.1063/1.1679012} {\bibfield  {journal}
  {\bibinfo  {journal} {J. Chem. Phys.}\ }\textbf {\bibinfo {volume} {58}},\
  \bibinfo {pages} {4496} (\bibinfo {year} {1973})}\BibitemShut {NoStop}%
\bibitem [{\citenamefont {Roothaan}(1951)}]{Roothaan1951}%
  \BibitemOpen
  \bibfield  {author} {\bibinfo {author} {\bibfnamefont {C.~C.~J.}\
  \bibnamefont {Roothaan}},\ }\href {\doibase 10.1103/RevModPhys.23.69}
  {\bibfield  {journal} {\bibinfo  {journal} {Rev. Mod. Phys.}\ }\textbf
  {\bibinfo {volume} {23}},\ \bibinfo {pages} {69} (\bibinfo {year}
  {1951})}\BibitemShut {NoStop}%
\bibitem [{Note6()}]{Note6}%
  \BibitemOpen
  \bibinfo {note} {The idea of selecting $ij$ like the \protect \textit {heat
  bath} excitation generator was communicated by Pablo L\'opez R\'ios (personal
  communication).}\BibitemShut {Stop}%
\bibitem [{\citenamefont {Dunning}(1989)}]{Dunning1989}%
  \BibitemOpen
  \bibfield  {author} {\bibinfo {author} {\bibfnamefont {T.~H.}\ \bibnamefont
  {Dunning}},\ }\href {\doibase 10.1063/1.456153} {\bibfield  {journal}
  {\bibinfo  {journal} {J. Chem. Phys.}\ }\textbf {\bibinfo {volume} {90}},\
  \bibinfo {pages} {1007} (\bibinfo {year} {1989})}\BibitemShut {NoStop}%
\bibitem [{\citenamefont {Vigor}\ \emph {et~al.}(2016)\citenamefont {Vigor},
  \citenamefont {Spencer}, \citenamefont {Bearpark},\ and\ \citenamefont
  {Thom}}]{Vigor2016}%
  \BibitemOpen
  \bibfield  {author} {\bibinfo {author} {\bibfnamefont {W.~A.}\ \bibnamefont
  {Vigor}}, \bibinfo {author} {\bibfnamefont {J.~S.}\ \bibnamefont {Spencer}},
  \bibinfo {author} {\bibfnamefont {M.~J.}\ \bibnamefont {Bearpark}}, \ and\
  \bibinfo {author} {\bibfnamefont {A.~J.~W.}\ \bibnamefont {Thom}},\ }\href
  {\doibase 10.1063/1.4943113} {\bibfield  {journal} {\bibinfo  {journal} {J.
  Chem. Phys.}\ }\textbf {\bibinfo {volume} {144}},\ \bibinfo {pages} {094110}
  (\bibinfo {year} {2016})},\ \Eprint {http://arxiv.org/abs/1601.00865}
  {arXiv:1601.00865} \BibitemShut {NoStop}%
\bibitem [{Note7()}]{Note7}%
  \BibitemOpen
  \bibinfo {note} {Feature implemented by Charles Scott.}\BibitemShut {Stop}%
\bibitem [{Note8()}]{Note8}%
  \BibitemOpen
  \bibinfo {note} {Personal Communication with Ali Alavi and Pablo L\'opez
  R\'ios. This is also implemented in NECI \protect \url
  {https://github.com/ghb24/NECI_STABLE}.}\BibitemShut {Stop}%
\end{thebibliography}
%merlin.mbs apsrev4-1.bst 2010-07-25 4.21a (PWD, AO, DPC) hacked
%Control: key (0)
%Control: author (8) initials jnrlst
%Control: editor formatted (1) identically to author
%Control: production of article title (-1) disabled
%Control: page (0) single
%Control: year (1) truncated
%Control: production of eprint (0) enabled
%

\end{document}